\documentclass[12pt]{spieman}  
\usepackage{amsmath,amsfonts,amssymb}
\usepackage{graphicx}
\usepackage{setspace}
\usepackage{tocloft}
\usepackage{lineno}

\usepackage{GB_packages}
\newacronym[type=symbolslist]{M}{\ensuremath{[\text{M}]}}{volume fraction of Melanosomes}
\newacronym[type=symbolslist]{dLpathAvg}{\ensuremath{\overline{\Delta{\langle L \rangle}}}}{average difference in total optical path-length}
\newacronym[type=symbolslist]{wph}{\ensuremath{w_\gamma}}{photon weight}
\newacronym[type=symbolslist]{V}{\ensuremath{V}}{voxel Volume}
\newacronym[type=symbolslist]{SpO2}{\ensuremath{\text{SpO}_2}}{pulsatile blood Oxygen Saturation}
\newacronym[type=symbolslist]{SaO2}{\ensuremath{\text{SaO}_2}}{arterial blood Oxygen Saturation}

\alwaysFloatBarrier

\usepackage{todonotes}

\title{Dual-ratio approach to pulse oximetry and the effect of skin tone}

\author[*]{Giles~Blaney}
\author[ ]{Jodee~Frias}
\author[ ]{Fatemeh~Tavakoli}
\author[ ]{Angelo~Sassaroli}
\author[ ]{Sergio~Fantini}
\affil[ ]{Tufts University, Department of Biomedical Engineering, Medford, MA USA, 02155}

\cftpagenumbersoff{figure}
\cftpagenumbersoff{table} 
\begin{document} 
\maketitle

\begin{abstract} 

\noindent\textbf{Significance:} 
Pulsatile blood Oxygen Saturation (SpO\textsubscript{2}) via pulse oximetry is a valuable clinical metric for assessing oxygen delivery. Individual anatomical features, including skin tone, may affect current optical pulse oximetry methods.

\noindent\textbf{Aim:} 
Develop an optical pulse oximetry method based on Dual-Ratio (DR) measurements to suppress individual anatomical features on SpO\textsubscript{2}.

\noindent\textbf{Approach:} 
Design a DR-based finger pulse oximeter, hypothesizing that DR would suppress confounds from optical coupling and superficial tissue-absorption. This method is tested using Monte Carlo (MC) simulations and \textit{in vivo} experiments.

\noindent\textbf{Results:} 
Different melanosome volume fraction in the epidermis, a surrogate for skin tone, cause changes in the recovered SpO\textsubscript{2} on the order of 1\%. Different heterogeneous pulsatile hemodynamics cause greater changes on the order of 10\%. SpO\textsubscript{2} recovered with DR measurements showed less variability than the traditional Single-Distance (SD) transmission method.

\noindent\textbf{Conclusions:} 
For the models and methods considered here, SpO\textsubscript{2} measurements are more strongly impacted by heterogeneous pulsatile hemodynamics than by melanosome volume fraction. This is consistent with previous reports that, the skin tone bias is smaller than the observed variation in recovered SpO\textsubscript{2} across individual people. The partial suppression of variability in the SpO\textsubscript{2} recovered by DR suggests promise of DR for pulse oximetry.
\end{abstract}

\keywords{optical pulse oximetry, blood oxygen saturation, near-infrared spectroscopy, melanin, hemodynamics, dual-ratio}

{\noindent \footnotesize\textbf{*}Giles~Blaney, Ph.D.,  \linkable{Giles.Blaney@tufts.edu} }

\begin{spacing}{1} 

\section{Introduction}\label{sect:intro} 
Pulse oximetry allows for the non-invasive measurement of \gls{SpO2} \ie{a surrogate for \gls{SaO2}} in various clinical settings.\cite{Tremper_Chest89_PulseOximetry, Severinghaus_Anesth.Analg.07_TakuoAoyagi, Nitzan_MDER14_PulseOximetry, Chan_RespiratoryMedicine13_PulseOximetry, Leppanen_Advancesinthediagnosisandtreatmentofsleepapnea:Fillingthegapbetweenphysiciansandengineers22_PulseOximetry}
\gls{SpO2} measurements by pulse oximetry have become ubiquitous in modern healthcare, providing valuable real-time assessment of patients' oxygen delivery.
The history of pulse oximetry may be considered to start with Glenn~Millikan, who invented the first practical oximeter in the 1940s.\cite{Millikan_RSI42_OximeterInstrument,Tremper_Chest89_PulseOximetry}
This invention was followed by Takuo Aoyagi's next technological advance in the 1970s when they developed pulse oximetry into something similar to today's technology.\cite{Severinghaus_Anesth.Analg.07_TakuoAoyagi}
However, despite pulse oximetry's widespread adoption and long history, open questions still exist regarding how differences between different people, such as skin tone, would confound the recovered \gls{SpO2}.\cite{Swamy_MBEC24_PulseOximeter,Setchfield_JBO24_EffectSkin, Al-Halawani_Physiol.Meas.23_ReviewEffect, Shi_BMCMed.22_AccuracyPulse, Cabanas_Sensors22_SkinPigmentation, Bierman_Br.J.Anaesth.24_MelaninBias, Martin_Br.J.Anaesth.24_EffectSkin, Mantri_Biomed.Opt.Express22_ImpactSkin, Moradi_Des.Qual.Biomed.Technol.XVII24_ModelingLighttissue}
These questions open the door for modern investigations of the technique and the proposal of novel oximetry methods.
\par

Various recent publications have focused on the impact of skin tone on \gls{SpO2} readings by pulse oximetry.
The measurement relies on the ratio of pulsatile optical signals at red and infrared wavelengths.
Therefore, the spectral dependence of melanin absorption may result in different confounds at different wavelengths, leading to an error in \gls{SpO2} when the wrong assumption about skin tone is made.
A recent letter examining the measured \gls{SpO2} versus \gls{SaO2} on a large population of patients found a positive bias in \gls{SpO2} for Black versus White patients \ie{the true \gls{SaO2} was lower on average for Black patients compared to White who showed the same \gls{SpO2} reading}.\cite{SjodingMichaelW._N.Engl.J.Med.20_RacialBias}
Furthermore, other recent studies, reviews, and meta-analyses suggest a similar bias.\cite{Al-Halawani_Physiol.Meas.23_ReviewEffect, Shi_BMCMed.22_AccuracyPulse, Cabanas_Sensors22_SkinPigmentation, Martin_Br.J.Anaesth.24_EffectSkin}
In these cases, the bias is on the order of a few percent on \gls{SpO2}, but this bias may become more pronounced at lower \gls{SaO2}.\cite{Swamy_MBEC24_PulseOximeter}
\par

Further recent studies seek to understand the possible optical origins of these biases and investigate ways to mitigate them by modeling the pulse oximetry measurements using methods such as \gls{MC} simulations.
One such study examined potential confounds affecting the recovered \gls{SpO2} and found that the modeled \gls{M} had a slight effect, and its primarily impact was on the \gls{Lpath} and the measured \gls{I}.\cite{Chatterjee_Sensors19_MonteCarlo}
A second study utilized similar methods but focused more directly on the effect of \gls{M} on \gls{SpO2} calibration, finding a positive bias for darker skin tones around \SIrange{1}{2}{\percent}.\cite{Moradi_Des.Qual.Biomed.Technol.XVII24_ModelingLighttissue}
Therefore, the biomedical optics field is actively investigating the observed skin tone bias in \gls{SpO2} measurements, but a definite consensus has yet to be reached.
The lack of definite consensus may suggest that there are differences between persons with different skin tones that still need to be fully captured in the current models.
Furthermore, new pulse oximetry methods have yet to be proposed that will reduce this bias in the optical measurement itself.
These novel methods will be needed so that pulse oximetry can be a valid clinical tool for all patients regardless of skin tone.
\par

In this work, we contribute to the investigation of the effect of skin tone on \gls{SpO2} measurements and propose a novel pulse oximetry method based on the \gls{DR} technique.
Our discussion is based on \gls{MC} simulations and \invivo{} measurements of \gls{SpO2} on healthy human subjects.
The \gls{MC} model allows us to investigate the effect of the \acrfull{M} in the epidermis on pulse oximetry measurements given either homogenous or heterogeneous pulsatile hemodynamics.
The novel component is the \gls{DR} measurement type applied to a human finger.
This geometry was inspired by a \gls{DR} method we developed for measuring absolute optical properties of turbid media in a cuvette.\cite{Blaney_App.Sci.22_MethodMeasuring}
\Gls{DR}\cite{Blaney_JBO23_DualratioApproach, Blaney_App.Sci.22_MethodMeasuring} is a measurement type based on the previously developed \gls{DS}\cite{Blaney_JBio20_PhaseDualslopes, Sassaroli_JOSAA19_DualslopeMethod} and \gls{SC}\cite{Hueber_Opt.Tomogr.Spectrosc.TissueIII99_NewOptical} techniques.
These techniques are insensitive to coupling changes and have suppressed \gls{sen} near the optodes.
These features of \gls{DR} have the potential to make \gls{DR} advantageous for \gls{SpO2} measurements as a result of the small sensitivity to skin-to-optode coupling or dynamics present in superficial tissue \ie{the epidermis}.
Therefore, in the following sections, we combine \gls{MC} models with \invivo{} data to investigate \gls{M}- or pulsatile-hemodynamic-heterogeneity- based confounds on \gls{SpO2} and the extent to which those confounds affect either the traditional \gls{SD} measurements in transmittance or the novel \gls{DR} measurement type applied to finger pulse oximetry.
\par

\section{Methods}\label{sect:meth} 

Methods for this work can be divided into two categories: \gls{MC} simulations (\autoref{sect:meth:MC}) and \invivo{} experimental measurements (\autoref{sect:meth:exp}). 
Outputs from the \gls{MC} simulations were used to analyze and interpret the experimental \invivo{} data.
We start by describing the measurement geometry (\autoref{sect:meth:geo}) and measurement types (\autoref{sect:meth:measTyps}) which are common to both.
\par

\subsection{Measurement geometry}\label{sect:meth:geo} 

\begin{figure}[bht]
	\begin{center}
		\includegraphics{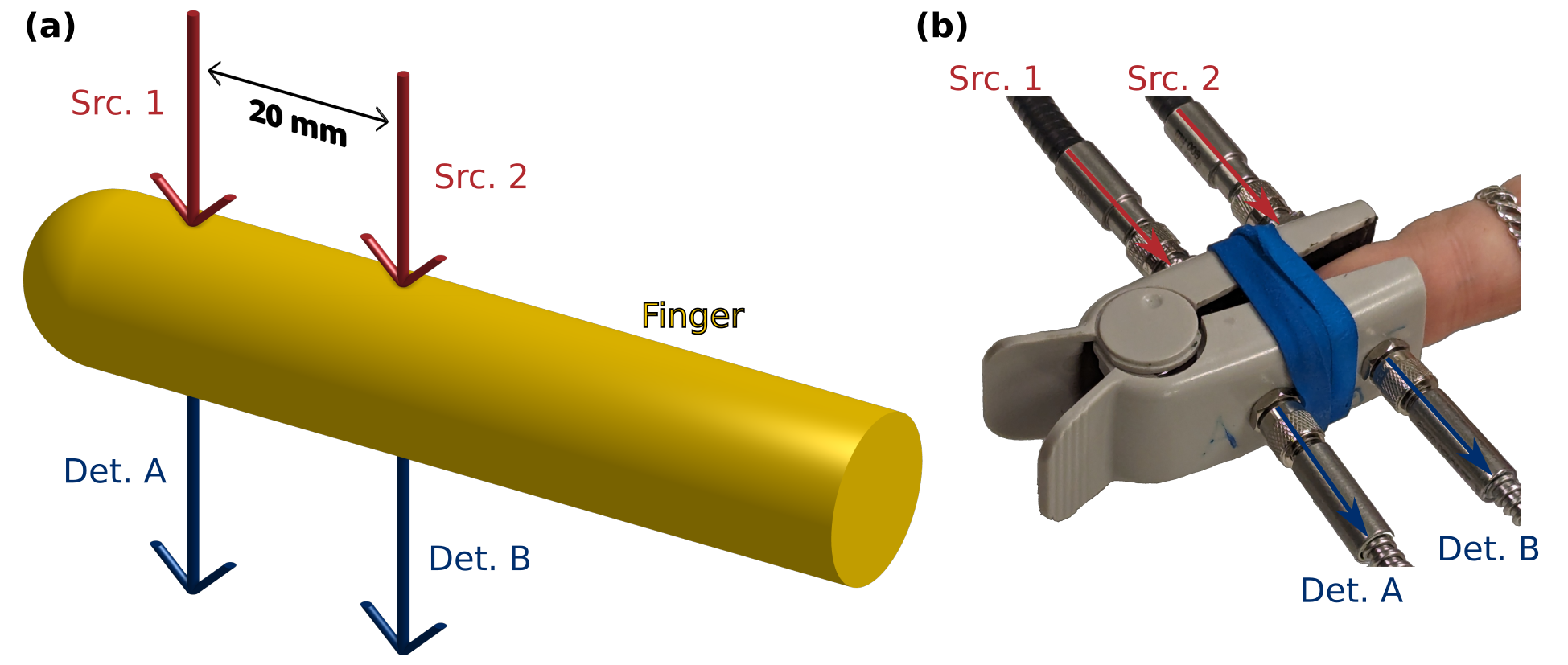}
	\end{center}
	\caption 
	{\label{fig:basicGeo}
		{(a)} Schematic of the measurement geometry. 
		Two Sources (Src.; 1 \& 2) and two Detectors (Det.; A \& B) were utilized in a transmission geometry through the finger to achieve a \acrfull{DR} set. 
		Data were collected from all four possible \acrfull{SD} source-detector pairs \ie{1A, 1B, 2A, 2B}.
		Src.~1/Det.~A and Src.~2/Det.~B were spaced \SI{20}{\milli\meter} apart. 
		{(b)} Photo of the probe which utilized four optical fibers \ie{one for each optode} and a standard pulse-oximeter finger clip.
		The finger was oriented so the sources were on the nail/knuckle side, and Src.~1 was placed behind the nail such that light did not enter the finger through the nail.} 
\end{figure} 

We considered the geometry in \autoref{fig:basicGeo} for all measurements and simulations in this work.
Two source locations and two detector locations were used in a transmission geometry.
Sources were named using numbers (1 \& 2) while detectors were named using letters (A \& B).
As shown in \autoref{fig:basicGeo}{(a)}, detector~A was placed in-line \ie{in transmission} with source~1 and, similarly, detector~B was in-line with source~2.
The spacing between sources~1~and~2 was \SI{20}{\milli\meter}, as shown in \autoref{fig:basicGeo}{(a)}.
Since corresponding sources and detectors were in-line, detectors~A~and~B were also spaced by the same \SI{20}{\milli\meter}.
\autoref{fig:basicGeo}{(b)} is a photo of the real-life measurement setup which realizes \autoref{fig:basicGeo}{(a)} using optical fibers and a modified pulse oximetry finger clip.
\par

\subsection{Measurement types}\label{sect:meth:measTyps} 

Two types of measurement are considered in this work, \gls{SD} and \gls{DR}.
\Gls{SD} is the traditional measurement utilized in \gls{NIRS} which is based upon the changes \ie{with respect to baseline} of the natural logarithm of \gls{I} measured between one source and one detector.
These \gls{SD} data may be converted to an effective \gls{dmua} using the \gls{Lpath}\footnote{If one wishes to consider the \gls{rho} and the \gls{DPF} instead, we can write $\as{Lpath}=\as{rho}\as{DPF}$.} according to the following equation:
\begin{equation}\label{equ:SDdmua}
	\as{dmua}_{,\text{\as{SD}}}=-\frac{\ln\left[\as{I}\right]-\ln\left[\as{I}_0\right]}{\as{Lpath}}
\end{equation}
where $\as{I}_0$ is the baseline \acrlong{I} and the subscript \as{SD} signifies that the quantity is an effective recovered \gls{dmua} from the \gls{SD} measurement type. 
For the measurement geometry in this work (\autoref{fig:basicGeo}), there are four \glspl{SD} \ie{1A, 1B, 2A, \& 2B}, however, for the results in this work, we focus on 1A.
\par

The second measurement type considered here is \gls{DR}.\cite{Blaney_JBO23_DualratioApproach, Blaney_App.Sci.22_MethodMeasuring}
\Gls{DR} is defined as the geometric mean of the ratio between \gls{I} measurements at a long and short source-detector distance.
Changes in the natural logarithm of this geometric mean can be converted to \gls{dmua} using a similar form as \autoref{equ:SDdmua}:
\begin{equation}\label{equ:DRdmua}
	\as{dmua}_{,\text{\as{DR}}}=-
		\frac{\ln\left[\sqrt{\frac{I_\text{1B}I_\text{2A}}{I_\text{1A}I_\text{2B}}}\right]-
		\ln\left[\sqrt{\frac{I_{\text{1B},0}I_{\text{2A},0}}{I_{\text{1A},0}I_{\text{2B},0}}}\right]}
		{\as{dLpathAvg}}
\end{equation}
where the subscript \as{DR} signifies that the quantity is an effective recovered \gls{dmua} from the \gls{DR} measurement type. 
In this case \ie{\autoref{equ:DRdmua}}, the proportionality constant is the negative inverse of the \acrfull{dLpathAvg} between short and long source-detector distances instead of the negative inverse of \gls{Lpath} as in \autoref{equ:SDdmua}.
This, \gls{dLpathAvg} will be defined later in \autoref{equ:dLpathAvg}.
\par

In either case, a measurement of \gls{dmua} at two or more wavelengths can be converted to a \gls{dHbO2} and a \gls{dHb} using Beer's law and their known extinction coefficients.\cite{Prahl_98_TabulatedMolar}
For this work, we considered four wavelengths with values of \SIlist{690;730;800;830}{\nano\meter}.
\par

\subsection{Monte-Carlo model and simulations}\label{sect:meth:MC} 

The \gls{MC} model and simulations in this work used the voxel based Monte-Carlo~eXtreme (MCX; rev0313d4~v2020)\cite{Fang_OE09_MonteCarlo} called from MATrix~LABoratory  (MATLAB; rev9.14.0.2286388~v2023a).
These simulations were run on a desktop computer with Linux~Mint~21.1, an AMD~Ryzen~9 7950X3D, 128~GB of main memory, and a NVIDIA~GeForce~RTX~4090 with 24~GB of graphics memory.
\par

Two types of \gls{MC} simulations were run for two different purposes. 
The first utilized a coarse voxel size and focused on detecting photons at the detector positions to determine the \gls{lpath_i}, where the different regions ($i$s) were associated with different tissue types.
This \gls{MC} to find \glspl{lpath_i} launched \num{e9}~photons and used a voxel size of \SI{0.25x0.25x0.25}{\milli\meter}.
The second \gls{MC} type utilized a fine voxel size with the goal of generating a high resolution \gls{PHI} spatial distribution.
This \gls{MC} type launched \num{10e9}~photons and used a voxel size of \SI{25x25x25}{\micro\meter}.
The two \gls{MC} types utilized a time range of \SIrange{0}{10}{\nano\second} with only one time bin so that the results are representative of \gls{CW} methods.
Additionally, all \gls{MC} simulations were run three separate times with different random seeds to determine the repeatability of the results.
\par

\subsubsection{Finger model} 

\begin{figure}[bht]
	\begin{center}
		\includegraphics{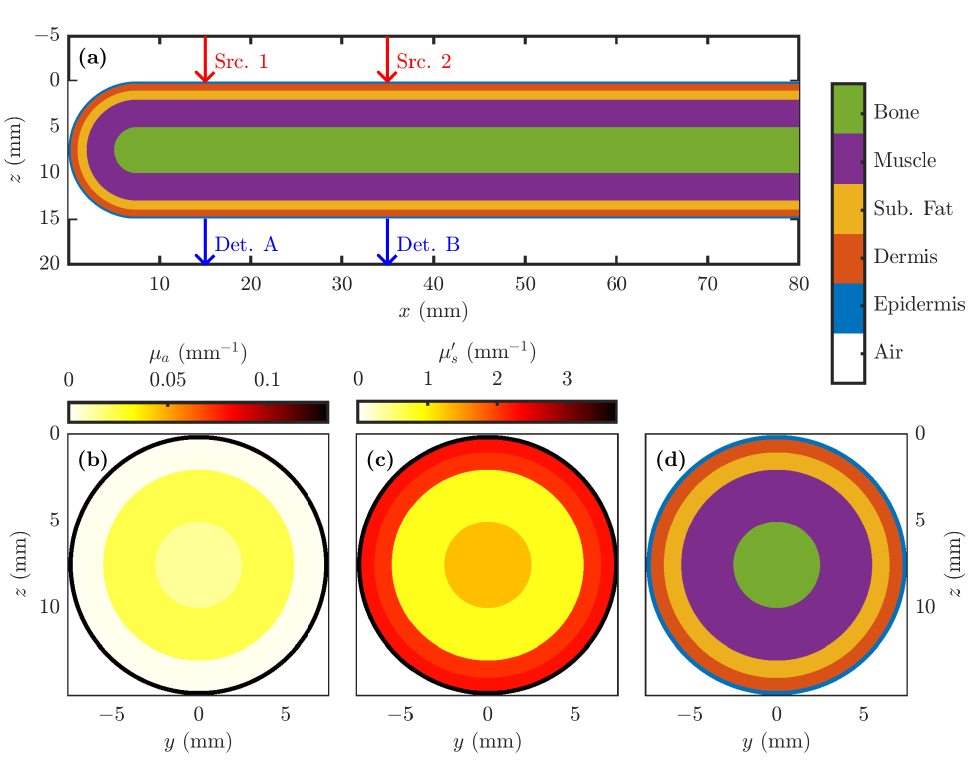}
	\end{center}
	\caption 
	{\label{fig:MCgeo}
		\Acrfull{MC} finger model. 
		The model consists of a cylinder \SI{80}{\milli\meter} long and \SI{15}{\milli\meter} in diameter with a \SI{15}{\milli\meter} diameter hemisphere representing the finger tip. Source~(Src.)~1 is placed at $(15\hat{x})$~\si{\milli\meter}, Src.~2 at $(35\hat{x})$~\si{\milli\meter}, Detector~(Det.)~A at $(15\hat{x}+15\hat{z})$~\si{\milli\meter}, and Det.~B at $(35\hat{x}+15\hat{z})$~\si{\milli\meter}.
		{(a)} $xz$ slice at $y=0$~\si{\milli\meter} of the finger model. 
		{(b)-(d)} $yz$ slice at $x=40$~\si{\milli\meter}.
		{(b)} Map of the \acrfull{mua} for the lowest \acrfull{M} case ($\as{M}=0.013$) and the \SI{800}{\nano\meter} wavelength.
		{(c)} Map of the \acrfull{musp} for the \SI{800}{\nano\meter} wavelength.
		{(d)} Slice of the finger model with the same color-scale as (a).
	}
\end{figure}

\begin{table}[bht]
	\caption{Modeled optical properties and chromophore concentrations in various tissue regions} 
	\label{tab:optProps}
	\begin{center}       
		\begin{tabular}{lS||S[table-format=1.4]S[table-format=1.1]S[table-format=1.2]S[table-format=1.2]|cccc|r}
			\multirow{5}{1cm}{Tissue} & 
			{\multirow{5}{0.8cm}{\centering\as{lam} (\si{\nano\meter})}} & 
			{\multirow{5}{1.1cm}{\centering\as{mua} (\si{\per\milli\meter})}} & 
			{\multirow{5}{1.1cm}{\centering\as{musp} (\si{\per\milli\meter})}} & 
			{\multirow{5}{0.8cm}{\centering\as{g}}} & 
			{\multirow{5}{0.8cm}{\centering\as{n}}} & 
			{\multirow{5}{0.8cm}{\centering\rotatebox{90}{\as{HbO2} (\si{\micro\Molar})}}} & 
			{\multirow{5}{0.8cm}{\centering\rotatebox{90}{\as{Hb} (\si{\micro\Molar})}}} & 
			{\multirow{5}{0.8cm}{\centering\rotatebox{90}{\as{Wfrac} (\si{\liter\per\liter_{tis}})}}} & 
			{\multirow{5}{0.8cm}{\centering\rotatebox{90}{\as{Lfrac} (\si{\liter\per\liter_{tis}})}}} & 
			{\multirow{5}{0.8cm}{\centering$\Delta r$\textsuperscript{\textdagger} (\si{\milli\meter})}} \\ 
			&&&&&&&&&&\\
			&&&&&&&&&&\\
			&&&&&&&&&&\\
			&&&&&&&&&&\\
			\hline\hline\hline
			
			\multirow{4}{*}{Epidermis\cite{Jacques_PMB13_OpticalProperties,Jacques_PMB13_ErratumOptical,Salomatina_JBO06_OpticalProperties,Ma_OL05_BulkOptical,Warner_JID88_ElectronProbe,Schiebener_JPCRef.Dat.90_RefractiveIndex}}
			& \num{690} & {\textdaggerdbl} & 4.3 & 0.90 & 1.48 &
			\multirow{4}{*}{0.0} & \multirow{4}{*}{0.0} & \multirow{4}{*}{0.20} & \multirow{4}{*}{0.00} & \multirow{4}{*}{0.25} \\
			& \num{730} & {\textdaggerdbl} & 4.1 & 0.91 & 1.48 &&&&&\\
			& \num{800} & {\textdaggerdbl} & 3.7 & 0.92 & 1.48 &&&&&\\
			& \num{830} & {\textdaggerdbl} & 3.6 & 0.92 & 1.48 &&&&&\\
			
			\hline
			\multirow{4}{*}{Dermis\cite{Jacques_PMB13_OpticalProperties,Jacques_PMB13_ErratumOptical,Salomatina_JBO06_OpticalProperties,Ma_OL05_BulkOptical,Choudhury_PhotonicTher.Diagn.VI10_LinkingVisual,Schiebener_JPCRef.Dat.90_RefractiveIndex}}
			& \num{690} & 0.0018 & 2.6 & 0.90 & 1.39 &
			\multirow{4}{*}{1.8} & \multirow{4}{*}{2.9} & \multirow{4}{*}{0.65} & \multirow{4}{*}{0.00} & \multirow{4}{*}{0.75} \\
			& \num{730} & 0.0022 & 2.5 & 0.91 & 1.39 &&&&&\\
			& \num{800} & 0.0023 & 2.2 & 0.92 & 1.39 &&&&&\\
			& \num{830} & 0.0029 & 2.2 & 0.92 & 1.39 &&&&&\\
			
			\hline
			\multirow{4}{*}{Sub.~Fat\cite{Jacques_PMB13_OpticalProperties,Jacques_PMB13_ErratumOptical,Salomatina_JBO06_OpticalProperties,Peters_PMB90_OpticalProperties,Jakubowski_JBO04_MonitoringNeoadjuvant,Schiebener_JPCRef.Dat.90_RefractiveIndex}}
			& \num{690} & 0.0023 & 2.4 & 0.98 & 1.49 &
			\multirow{4}{*}{9.5} & \multirow{4}{*}{3.0} & \multirow{4}{*}{0.11} & \multirow{4}{*}{0.69} & \multirow{4}{*}{1.00} \\
			& \num{730} & 0.0022 & 2.2 & 0.98 & 1.49 &&&&&\\
			& \num{800} & 0.0028 & 2.1 & 0.98 & 1.49 &&&&&\\
			& \num{830} & 0.0035 & 2.0 & 0.98 & 1.49 &&&&&\\
			
			\hline
			\multirow{4}{*}{Muscle\cite{Jacques_PMB13_OpticalProperties,Jacques_PMB13_ErratumOptical,Venkata_12_DeterminationOptical,Matcher_AO97_VivoMeasurements,Mitchell_JBC45_CHEMICALCOMPOSITION,Schiebener_JPCRef.Dat.90_RefractiveIndex}}
			& \num{690} & 0.025 & 1.0 & 0.95 & 1.37 &
			\multirow{4}{*}{75} & \multirow{4}{*}{42} & \multirow{4}{*}{0.80} & \multirow{4}{*}{0.00} & \multirow{4}{*}{3.00} \\
			& \num{730} & 0.019 & 0.9 & 0.95 & 1.36 &&&&&\\
			& \num{800} & 0.023 & 0.8 & 0.95 & 1.36 &&&&&\\
			& \num{830} & 0.026 & 0.8 & 0.95 & 1.36 &&&&&\\
			
			\hline
			\multirow{4}{*}{Bone\cite{Jacques_PMB13_OpticalProperties,Jacques_PMB13_ErratumOptical,Bevilacqua_AO00_BroadbandAbsorption,Firbank_PMB93_MeasurementOptical,Alexandrakis_PMB05_TomographicBioluminescence,Weaver_CBPA89_TissueBlood,Harrison_Blood02_OxygenSaturation,Mitchell_JBC45_CHEMICALCOMPOSITION,Schiebener_JPCRef.Dat.90_RefractiveIndex}}
			& \num{690} & 0.0082 & 1.4 & 0.94 & 1.45 &
			\multirow{4}{*}{61} & \multirow{4}{*}{8.8} & \multirow{4}{*}{0.32} & \multirow{4}{*}{0.00} & \multirow{4}{*}{2.50} \\
			& \num{730} & 0.0083 & 1.3 & 0.94 & 1.45 &&&&&\\
			& \num{800} & 0.014 & 1.3 & 0.94 & 1.45 &&&&&\\
			& \num{830} & 0.016 & 1.3 & 0.94 & 1.45 &&&&&\\
		\end{tabular}
	\end{center}
	\small Symbols: \Acrfull{lam}, \acrfull{mua}, \acrfull{musp}, \acrfull{g}, \acrfull{n}, \acrfull{HbO2}, \acrfull{Hb}, \acrfull{Wfrac}, \acrfull{Lfrac} \\
	Note \textdagger: For bone $\Delta r$ represents the radius, for other tissues $\Delta r$ is the radial thickness. \\
	Note \textdaggerdbl: See \autoref{tab:epiMUA}.
\end{table} 

\begin{table}[bht]
	\caption{Modeled \acrfull{mua} for various \acrfull{M}\cite{Jacques_PMB13_OpticalProperties,Jacques_PMB13_ErratumOptical,Jacques_98_MelanosomeAbsorption,Jacques_18_ExtinctionCoefficient,Jacques_18_OpticalAbsorption,Jacques_PP91_MelanosomeThreshold,Choudhury_PhotonicTher.Diagn.VI10_LinkingVisual,Warner_JID88_ElectronProbe}} 
	\label{tab:epiMUA}
	\begin{center}       
		\begin{tabular}{l|cccc} 
			\multirow{2}{0.8cm}{\centering\as{lam} (\si{\nano\meter})} & \multicolumn{4}{c}{\as{mua} (\si{\per\milli\meter})} \\
			& $\as{M}=0.013$ & $\as{M}=0.152$ & $\as{M}=0.291$& $\as{M}=0.430$ \\
			\hline\hline\hline
			\num{690} & 0.22 & 2.6 & 4.9 & 7.2 \\
			\num{730} & 0.18 & 2.1 & 4.0 & 5.9 \\
			\num{800} & 0.13 & 1.5 & 2.9 & 4.3 \\
			\num{830} & 0.12 & 1.3 & 2.6 & 3.8 \\
		\end{tabular}
	\end{center}
	\small Symbols: \Acrfull{lam}, \acrfull{mua}, \acrfull{M} \\
\end{table}

The finger was modeled as an \SI{80}{\milli\meter} long and \SI{15}{\milli\meter} diameter cylinder with a hemisphere at one end to represent the finger-tip (\autoref{fig:basicGeo}{(a)} and \autoref{fig:MCgeo}).
As shown in \autoref{fig:MCgeo}, the coordinate system considered the $x$-axis along the length of the cylinder, the $y$-axis along the width, and the $z$-axis along the height \ie{nail to pad}.
The origin was placed with $x=\SI{0}{\milli\meter}$ at the tip of the hemisphere \ie{the finger-tip}, $y=\SI{0}{\milli\meter}$ at the center of the cylinder, and $z=\SI{0}{\milli\meter}$ at the top of the cylinder \ie{the plane that the sources are incident upon}. 
\par

Given this coordinate system, source~1 was placed at $(15\hat{x})$~\si{\milli\meter}, source~2 at $(35\hat{x})$~\si{\milli\meter}, detector~A at $(15\hat{x}+15\hat{z})$~\si{\milli\meter}, and detector~B at $(35\hat{x}+15\hat{z})$~\si{\milli\meter} (\autoref{fig:MCgeo}). 
For the first \gls{MC} type to find \glspl{lpath_i}, the detectors had radii of \SI{1.5}{\milli\meter} so that the detection area was the surface defined by the intersection of a sphere of this radius centered at the detector coordinate and the surface of the cylinder.
\par

Five tissue types were modeled as shells concentric with both the cylinder and hemisphere allowing each tissue type to be characterized by a layer thickness ($\Delta r$; \autoref{tab:optProps}). 
These tissue types were: epidermis, dermis, subcutaneous-fat (sub.~fat), muscle, and bone. 
Each tissue type was modeled with unique wavelength dependent optical properties, as described in \autoref{tab:optProps} and \autoref{tab:epiMUA}.
The \gls{MC} simulations were run for each of the four wavelengths \ie{\SIlist{690;730;800;830}{\nano\meter}}.
Furthermore, a range of \gls{M} was also considered (\autoref{tab:epiMUA}).
The first \gls{MC} type to find \gls{lpath_i}, considered \gls{M} in the range of \numrange{0.013}{0.430} linearly spaced over \num{1000} values.
The second \gls{MC} to find \gls{PHI}, considered four values for \gls{M} of \numlist{0.013;0.152;0.291;0.430}.
For the most part, chromophore concentrations and optical properties were assigned according to References~\citenum{Jacques_PMB13_OpticalProperties}~\&~\citenum{Jacques_PMB13_ErratumOptical}.
\autoref{tab:optProps} and \autoref{tab:epiMUA} specify each specific reference pertinent to each tissue type.
\par

\subsubsection{Calculation of path-lengths and sensitivities}\label{sect:meth:MC:LandSEN} 

\paragraph{Monte-Carlo for partial path-lengths}\label{sect:meth:MC:LandSEN:L}
The first \gls{MC} type was run using a coarse voxel size with the goal to determining the \acrfull{lpath_i} \ie{epidermis, dermis, subcutaneous-fat, muscle, and bone} and the \acrfull{Lpath} for each source-detector pair.\footnote{A separate \gls{MC} was run for each source and wavelength, such that \glspl{lpath_i} and \glspl{Lpath} were found for each source-detector pair \ie{1A, 1B, 2A, \& 2B} and each wavelength \ie{\SIlist{690;730;800;830}{\nano\meter}}. These \glspl{MC} were repeated three times to determine uncertainties in the results.}
This \gls{MC} was run white \ie{with zero \gls{mua}} and the \gls{mua} of the various tissue types was applied post-runtime.
For each detector \ie{A or B}, the \gls{lpath1} spent in each type of tissue by each detected photon was saved.
For a particular source-detector pair and wavelength, these \glspl{lpath1} were indexed by tissue region ($i$) and photon number ($\gamma$) so that we have $\as{lpath1}_{i,\gamma}$.
\par

To take into account the \gls{mua} of the different tissue regions ($\as{mua}_{,i}$) post-runtime, each individual \gls{wph} was calculated. 
This was done by scaling the photon \gls{wph} using Beer-Lambert law as follows:
\begin{equation}
	\as{wph}=\prod_i^{N_{\text{regions}}} e^{-\as{mua}_{,i} \as{lpath1}_{i,\gamma}}
\end{equation}
Then these \acrfullpl{wph} were used to calculate the weighted average of \glspl{lpath1} and yield the \acrfull{lpath_i} and the \acrfull{Lpath} as follows:
\begin{equation}
	\as{lpath_i}=\frac{\sum_{\gamma}^{N_{\text{detected}}} \as{wph} \as{lpath1}_{i,\gamma}}
		{\sum_{\gamma}^{N_{\text{detected}}} \as{wph}}
\end{equation}
\begin{equation}\label{equ:l2Lpath}
	\as{Lpath}=\sum_i^{N_{\text{regions}}} \as{lpath_i}
\end{equation}
where the $i$ subscript on \gls{lpath_i} represents the tissue region.
Unique values of \gls{lpath_i} and \gls{Lpath} were found for each source-detector pair and each wavelength.
\par

\Gls{Lpath} is the needed proportionality constant in \autoref{equ:SDdmua} to convert \gls{SD} \gls{I} data to \gls{dmua}.
For \gls{DR} we need the \gls{dLpathAvg} in \autoref{equ:DRdmua} to convert \gls{DR} \gls{I} data to \gls{dmua}.
To find \gls{dLpathAvg} we consider the \gls{Lpath} for each source-detector pair \ie{1A, 1B, 2A, \& 2B}, and \gls{dLpathAvg} is calculated as follows:
\begin{equation}\label{equ:dLpathAvg}
	\as{dLpathAvg}=\frac{\left(\as{Lpath}_\text{1B}-\as{Lpath}_\text{1A}\right)+\left(\as{Lpath}_\text{2A}-\as{Lpath}_\text{2B}\right)}{2}
\end{equation}
\par

Finally, the last output obtained from this first \gls{MC} type was the \gls{sen} for each tissue region.
In the \gls{SD} case, \gls{sen} for tissue region $i$ is calculated as a ratio of \gls{lpath_i} to \gls{Lpath} as follows:\cite{Blaney_JIOHS24_SpatialSensitivity,Blaney_22_EnablingDeep}
\begin{equation}\label{equ:senSD}
	\as{sen}_{\text{\as{SD}},i}=\frac{\as{lpath_i}}{\as{Lpath}}
\end{equation}
While for the \gls{DR} case, \gls{sen} for tissue region $i$ is given by the ratio of the average difference in $\as{lpath_i}$s to the average difference in \glspl{Lpath}:\cite{Blaney_JIOHS24_SpatialSensitivity,Blaney_22_EnablingDeep}
\begin{equation}\label{equ:senDR}
	\as{sen}_{\text{\as{DR}},i}=
	\frac{\left(\as{lpath_i}_{\text{1B}}-\as{lpath_i}_{\text{1A}}\right)+\left(\as{lpath_i}_{\text{2A}}-\as{lpath_i}_{\text{2B}}\right)}
	{\left(\as{Lpath}_\text{1B}-\as{Lpath}_\text{1A}\right)+\left(\as{Lpath}_\text{2A}-\as{Lpath}_\text{2B}\right)}
\end{equation}
In both the \gls{SD} and \gls{DR} cases, $\as{sen}_i$ is interpreted as the ratio of the recovered effective \gls{dmua} from a measurement \ie{from \autoref{equ:SDdmua} or \autoref{equ:DRdmua}} and a true local $\as{dmua}_{,i}$ in tissue region $i$. 
This concept is expressed by the following equation:\cite{Blaney_JIOHS24_SpatialSensitivity,Blaney_22_EnablingDeep}
\begin{equation}\label{equ:sen_dmua}
	\as{sen}_i=\frac{\as{dmua}_{,\text{recovered}}}{\as{dmua}_{,i}}
\end{equation}
Furthermore, these definitions lead to the following property of $\as{sen}_i$s:\cite{Blaney_JIOHS24_SpatialSensitivity,Blaney_22_EnablingDeep}
\begin{equation}
	\sum_{i}^{N_{\text{regions}}} \as{sen}_i = 1
\end{equation}
which means that a homogeneous \gls{dmua} \ie{$\as{dmua}_{,i}$s are equal regardless of $i$} will result in a measured effective recovered \gls{dmua} that us equal to the true homogeneous perturbation.
\par

\paragraph{Monte-Carlo for fluence rate distribution}
The second \gls{MC} type was run with a fine voxel size with the goal of determining the spatial distributions of the \gls{PHI} and then calculating a high-resolution spatial map of \gls{sen}.\footnote{The first \gls{MC} type only yielded \gls{sen} for the five tissue regions, while the second \gls{MC} type aims to find \gls{sen} for each voxel to create a spatial map.}
The outputs of this \gls{MC} type were the \acrfull{PHI} normalized by source power distributions for a pencil beam placed at each source or detector location.
The \gls{PHI} distribution from a pencil beam at the detector locations was found by mirroring the \gls{PHI} distributions from the source locations about the plane defined by $z=\SI{7.5}{\milli\meter}$.
This approach of mirroring to find \gls{PHI} from the detectors is possible due to the symmetry in the modeled geometry (\autoref{fig:MCgeo}).
\par

These \gls{PHI} distributions were used to find \gls{sen} distributions based on a method similar to the adjoint method.\cite{Yao_BOE18_DirectApproach}
First, to motivate the adjoint method we write the \gls{lpath_j} with \gls{V} in terms of a \gls{PHI} and \glspl{R}:\cite{Blaney_22_EnablingDeep}
\begin{equation}\label{equ:lpath}
	\as{lpath_j}=\frac{\as{PHI}\left[\as{r}_\text{src}\rightarrow\as{r}_j\right]\as{R}\left[\as{r}_j\rightarrow\as{r}_\text{det}\right]}{\as{R}\left[\as{r}_\text{src}\rightarrow\as{r}_\text{det}\right]} \as{V}
\end{equation}
The arguments of \gls{PHI} or \gls{R} in \autoref{equ:lpath} specify the \glspl{r} of the voxel field point ($j$), the source (src), or the detector (det).
The direction of light transport is specified by the arrows in the argument ($\rightarrow$).
Next we approximate the \glspl{R} with \glspl{PHI} and apply the reciprocity relation $\as{PHI}\left[\as{r}_j\rightarrow\as{r}_\text{det}\right]=\left(\as{n}^2_\text{det}/\as{n}^2_j\right)\as{PHI}\left[\as{r}_\text{det}\rightarrow\as{r}_j\right]$ which accounts for the \gls{n} at the detector\footnote{When applying the adjoint method the \gls{n} at the detector is the \gls{n} of the medium just below the detector since these voxels below the detector are used to determine the detected \gls{PHI}.} and voxel.\cite{Aronson_JOSAA97_RadiativeTransfer}
Considering that \gls{V}, $\as{R}\left[\as{r}_\text{src}\rightarrow\as{r}_\text{det}\right]$, and $\as{n}_\text{det}$ are constants that do not depend on voxel position we lump them together into the constant $\beta$ and rewrite an approximation of \autoref{equ:lpath}:
\begin{equation}\label{equ:lpath_adjoint1}
	\as{lpath_j}\approxeq\beta\frac{\as{PHI}\left[\as{r}_\text{src}\rightarrow\as{r}_j\right]\as{PHI}\left[\as{r}_\text{det}\rightarrow\as{r}_j\right]}{\as{n}^2_j}
\end{equation}
To find $\beta$, we can apply \autoref{equ:l2Lpath} and use the \gls{Lpath} found from the first \gls{MC} type:
\begin{equation}
	\beta=\frac{\as{Lpath}}{
		\sum_{j}^{N_\text{voxels}}\frac{\as{PHI}\left[\as{r}_\text{src}\rightarrow\as{r}_j\right]\as{PHI}\left[\as{r}_\text{det}\rightarrow\as{r}_j\right]}{\as{n}^2_j}
		}
\end{equation}
for each source-detector pair. 
After obtaining $\beta$ using the \gls{Lpath} found from the first \gls{MC} type, we can use \autoref{equ:lpath_adjoint1} to find an approximation of \gls{lpath_j} for each voxel.
Finally, \autoref{equ:senSD} and \autoref{equ:senDR} are used to find $\as{sen}_j$ for each voxel and measurement type to create high-resolution spatial maps of \gls{sen}. 
\par

\subsubsection{Simulation of pulsatile hemodynamics and recovered pulsatile saturation}\label{sect:meth:MC:simSpO2} 

Using the \acrfull{sen} for each tissue region, we can model hemodynamic oscillations \ie{from cardiac pulsation} in each tissue region and then simulate the associated recovered \gls{dmua}.
To this aim, we used phasor notation to represent hemodynamic oscillations with a given amplitude and phase at a given frequency \ie{the heart-rate in this case}.
These hemodynamic phasors for \gls{HbO2} and \gls{Hb} were modeled in each tissue.
Then these phasors were converted to \gls{mua} phasors in each tissue at each of the four wavelengths using Beer's law and known extinction coefficients.\cite{Prahl_98_TabulatedMolar, Blaney_22_EnablingDeep}
Next, the recovered \gls{mua} phasors were found by a linear combination of the \gls{mua} phasors in each tissue region weighed by the $\as{sen}_i$ in tissue region $i$ (\autoref{equ:sen_dmua}).\cite{Blaney_22_EnablingDeep}
These simulated recovered \gls{mua} phasors at four wavelengths are then converted to recovered \gls{HbO2} and \gls{Hb} phasors, again using Beer's law.\cite{Prahl_98_TabulatedMolar, Blaney_22_EnablingDeep}
Finally, the \gls{SpO2} is obtained from these \gls{HbO2} and \gls{Hb} phasors ($\widetilde{\as{HbO2}}$ \& $\widetilde{\as{Hb}}$) as follows:
\begin{equation}\label{equ:SpO2}
	\as{SpO2}=\frac{\left|\widetilde{\as{HbO2}}\right|}{\left|\widetilde{\as{HbO2}}\right|+\left|\widetilde{\as{Hb}}\right|}
\end{equation}
We observe that for \gls{SpO2} in \autoref{equ:SpO2} to be representative of the blood oxygen saturation of a volume oscillating vasculature compartment \ie{as with \gls{SaO2}}, the phasors $\widetilde{\as{HbO2}}$ and $\widetilde{\as{Hb}}$ must be in-phase with each-other.\cite{Fantini_NeuroImage14_DynamicModel}
If this is not the case, one needs to apply a correction to take into account the phase difference between $\widetilde{\as{HbO2}}$ and $\widetilde{\as{Hb}}$.\cite{Kainerstorfer_JBO16_OpticalOximetry}
\par

For the simulations, the recovered \gls{SpO2} represents what would be recovered given the modeled hemodynamic phasors in each tissue region.
This can be done for either measurement type \ie{\gls{SD} or \gls{DR}} and, in the case of \gls{SD}, for each source-detector pair \ie{1A, 1B, 2A, \& 2B}.
Note that \autoref{equ:SpO2} not only applies to simulations but also represents how we calculate \gls{SpO2} in general for this work, including for the \invivo{} data.
We further emphasize that this method of recovering \gls{SpO2} with \autoref{equ:SpO2} from \gls{HbO2} and \gls{Hb} phasors individually does not account for the phase relationship between \gls{HbO2} and \gls{Hb}.
\par

\subsection{\textit{In vivo} measurements}\label{sect:meth:exp} 

\subsubsection{Recovery of pulsatile saturation}\label{sect:meth:exp:spo2} 

For the \invivo{} measurements, we used the finger clip probe shown in \autoref{fig:basicGeo}{(b)} and collected the \acrfull{I} between each source and detector \ie{1A, 1B, 2A, \& 2B} and for each wavelength \ie{\SIlist{690;730;800;830}{\nano\meter}}.\footnote{The protocol for these measurements is described in \autoref{sect:meth:exp:prot}.}
The first step in the analysis of these \invivo{} data was to convert these measured \gls{I} to \gls{dmua} for each wavelength and measurement type \ie{\gls{SD}\footnote{For this work the \gls{SD} pair focused upon was 1A.} and \gls{DR}}.
This conversion was done using \autoref{equ:SDdmua} and \autoref{equ:DRdmua}, assuming the \glspl{Lpath} or \glspl{dLpathAvg} obtained from the first \gls{MC} type described in \autoref{sect:meth:MC:LandSEN:L}.
Built into this assumption are the finger's optical properties and geometry, including the value of \gls{M} in the epidermis.
Therefore, the \invivo{} data may be analyzed with different assumed values for \gls{M} ($\as{M}_{ass}$), to investigate the effect of such assumption.
\par

From the temporal traces of \gls{dmua} at the four wavelengths, the temporal \glspl{dHbO2} and \glspl{dHb} were found using Beer's law.\cite{Prahl_98_TabulatedMolar}
Working with \gls{dHbO2} from \gls{DR} data we next find the heart-frequency \ie{heart-rate}, which is then assumed to be the same for all measurement types within one dataset.
To find this frequency, the data were first de-trended so that the first and last temporal points took the value of \SI{0}{\micro\Molar}.
Then, a high-pass filter with a cutoff of \SI{50/60}{\hertz} \ie{\SI{0.83}{\hertz}} was applied to the signal.
Next, the filtered temporal data is transformed into the Fourier domain using a \gls{FFT} with a Nuttall Blackman-Harris window.\cite{Nuttall_IEEETrans.81_WindowsVery}
Considering only frequencies below \SI{2.5}{\hertz}, the peak in the Fourier domain with the highest amplitude was identified.
To find the peak centroid, we considered the peak extending from the first minimum below the frequency of maximum amplitude to the first minimum above the said maximum amplitude frequency point.
The centroid frequency was calculated as a weighted average frequency, weighted by the amplitude of each frequency point which comprised the peak; this centroid was taken as the heart-frequency.
\par

Knowing the heart-frequency, the temporal signals of \gls{dHbO2} and \gls{dHb} associated with each data-type are band-pass filtered about it.
The band-pass filter utilized a central frequency equal to the heart-frequency and a bandwidth of \SI{10}{\milli\hertz}.\footnote{This small bandwidth filter is needed to apply the Hilbert transform in the next step.}
At this point the signals are analyzed in two different ways: an \gls{FFT} to determine amplitudes and a Hilbert transform to determine phases.
For the amplitudes, an \gls{FFT} of the band-passed signals was taken and the amplitude was determined as the integral of this Fourier spectrum from the beginning to the end of the heart-frequency peak defined as before.
The integral bounds are larger than the band-pass bandwidth so that the entire peak is included in the interval.
We will write the amplitudes found in this way as $\left|\widetilde{\as{HbO2}}\right|$ and $\left|\widetilde{\as{Hb}}\right|$.
To determine the phase of \gls{dHbO2} and \gls{dHb} at the heart-frequency we employed the Hilbert transform on the band-passed signals.
The phase reference was considered to be the phase of the change in total-hemoglobin concentration \ie{$\as{dHbO2} + \as{dHb}$} measured by \gls{DR}.
We will write the phases found in this way as $\angle\widetilde{\as{HbO2}}$ and $\angle\widetilde{\as{Hb}}$.
\par

Finally, we utilized the amplitudes, $\left|\widetilde{\as{HbO2}}\right|$ and $\left|\widetilde{\as{Hb}}\right|$, to calculate \gls{SpO2} using \autoref{equ:SpO2}.
These recovered \gls{SpO2} values are obtained for each measurement type but also for different assumed values of \gls{M}.
Therefore, in this work, we investigate how assumed values of \gls{M} affect the recovered \gls{SpO2}.
Further, we will utilize the recovered $\angle\widetilde{\as{HbO2}}$ and $\angle\widetilde{\as{Hb}}$ to aid in discussing how phase differences between \gls{HbO2} and \gls{Hb} oscillations affect the recovered \gls{SpO2}.
\par

\subsubsection{Determination of skin tone}\label{sect:meth:exp:skin} 

\begin{figure}[bht]
	\begin{center}
		\includegraphics{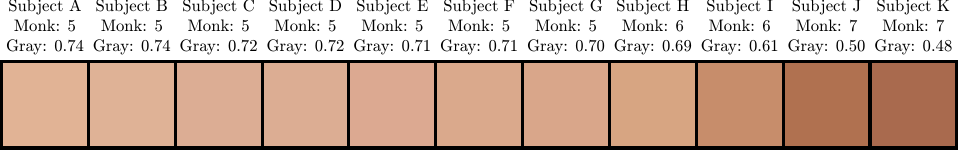}
	\end{center}
	\caption 
	{\label{fig:handSwatches}
		Color swatches of each subject's skin tone obtained from a photo of the back of the subject's hand. For each subject, the Monk scale\cite{Monk_2019} value and swatch gray-scale value (using Rec.ITU-R~BT.601-7) is also reported.
	} 
\end{figure}

To connect the values of \gls{M} in the \gls{MC} models with the \invivo{} data, we have quantified the skin tone of the human subjects we measured.
For the determination of skin tone we utilized the Monk scale.\cite{Monk_2019}
The Monk scale consists of \num{10} swatches which represent a wide range of skin tones, with Monk~1 being the lightest and Monk~10 being the darkest.
\par

To determine each subject's skin tone, a photo of the back of their hand was taken with a Canon~EOS~Rebel~T3i digital camera.
For these photos, the stock \SIrange{18}{55}{\milli\meter} lens was used set at the \SI{55}{\milli\meter} focal-length.
The camera took a photo from approximately \SI{1}{\meter} distance with a \SI{60}{\watt} incandescent light source at almost the same location as the camera.
The incandescent light source was the only light source in the room when the photo was taken.
The subject's hand was placed on a white background and the camera focused to the back of the hand.
Camera settings were set to full manual and were the same for each subject.
The primary camera settings were a ISO of \num{100}, an aperture of $f/5$, and an exposure time of \SI{0.3}{\second}.
We would like to note that these settings are considered to slightly overexpose the scene, however no part of the images were saturated.
\par

Each photo was saved in the Canon raw image format (CR2) and was \SI{5184x3456}{px} in size.
A crop of \SI{500x500}{px} around the center of the back of the subject's hand was used for further analysis.
This cropped image was averaged to find the average Red-Green-Blue (RGB) color value.
These color swatches of the RGB value for each subject are shown in \autoref{fig:handSwatches}.
At this point, we note that these RGB color values are comparable between each-other due to the control over camera setup and parameters, but the values would be difficult to compare to color values independently measured by other photographers.
\par

To quantify the subject's Monk scale value, we compared the subject's RGB skin-tone value to values derived from photos of Monk scale color swatches.
The Monk scale swatches were printed on white paper using a Canon~iR~ADV~C5250 color printer.
The printed swatch pallette was then photographed with the same settings and at the location as the photos of the subject's hand.
Finally, the determination of Monk scale value for each subject was found by minimizing the difference between the subject's RGB value and the RGB values for the Monk color swatches.
We again note that these values are comparable between subjects in this paper but would be difficult to compare to other photos or Monk scale values determined in different ways.
These Monk scale values are reported in \autoref{fig:handSwatches} and \autoref{tab:subInfo}.
\par

\subsubsection{Protocol and subjects} \label{sect:meth:exp:prot} 

\begin{table}[bht]
	\caption{Subject information} 
	\label{tab:subInfo}
	\begin{center}       
		\begin{tabular}{c|cccccc} 
			{\multirow{3}{1.2cm}{\centering Subject}} & 
			{\multirow{3}{0.8cm}{\centering Age}} & 
			{\multirow{3}{1.5cm}{\centering Sex at Birth}} & 
			{\multirow{3}{2.5cm}{\centering Monk Skin Tone\cite{Monk_2019}\\(See \autoref{fig:handSwatches})}} & 
			{\multirow{3}{1cm}{\centering Race}} & 
			{\multirow{3}{2cm}{\centering Ethnicity}} & 
			{\multirow{3}{3cm}{\centering Majority Ancestral Region}} \\
			&&&&&&\\
			&&&&&&\\
			\hline\hline\hline
			
			\multirow{2}{*}{A} & 
			\multirow{2}{*}{26} & 
			\multirow{2}{*}{Female} & 
			\multirow{2}{*}{5} & 
			\multirow{2}{*}{Multi} & 
			\multirow{2}{*}{Hispanic} & 
			\multirow{2}{3cm}{\centering North-America} \\
			&&&&&&\\
			\hline
			
			\multirow{2}{*}{B} & 
			\multirow{2}{*}{31} & 
			\multirow{2}{*}{Female} & 
			\multirow{2}{*}{5} & 
			\multirow{2}{*}{White} & 
			\multirow{2}{2cm}{\centering Non-Hispanic} & 
			\multirow{2}{3cm}{\centering Southern-Europe} \\
			&&&&&&\\
			\hline
			
			\multirow{2}{*}{C} & 
			\multirow{2}{*}{28} & 
			\multirow{2}{*}{Male} & 
			\multirow{2}{*}{5} & 
			\multirow{2}{*}{Asian} & 
			\multirow{2}{2cm}{\centering Non-Hispanic} & 
			\multirow{2}{3cm}{\centering Southeast-Asia} \\
			&&&&&&\\
			\hline
			
			\multirow{2}{*}{D} & 
			\multirow{2}{*}{30} & 
			\multirow{2}{*}{Male} & 
			\multirow{2}{*}{5} & 
			\multirow{2}{*}{White} & 
			\multirow{2}{2cm}{\centering Non-Hispanic} & 
			\multirow{2}{3cm}{\centering Northern- \& Eastern- Europe} \\
			&&&&&&\\
			\hline
			
			\multirow{2}{*}{E} & 
			\multirow{2}{*}{23} & 
			\multirow{2}{*}{Female} & 
			\multirow{2}{*}{5} & 
			\multirow{2}{*}{White} & 
			\multirow{2}{2cm}{\centering Non-Hispanic} & 
			\multirow{2}{3cm}{\centering Western-Europe} \\
			&&&&&&\\
			\hline
			
			\multirow{2}{*}{F} & 
			\multirow{2}{*}{25} & 
			\multirow{2}{*}{Male} & 
			\multirow{2}{*}{5} & 
			\multirow{2}{*}{Asian} & 
			\multirow{2}{2cm}{\centering Non-Hispanic} & 
			\multirow{2}{3cm}{\centering East-Asia} \\
			&&&&&&\\
			\hline
			
			\multirow{2}{*}{G} & 
			\multirow{2}{*}{59} & 
			\multirow{2}{*}{Male} & 
			\multirow{2}{*}{5} & 
			\multirow{2}{*}{White} & 
			\multirow{2}{2cm}{\centering Non-Hispanic} & 
			\multirow{2}{3cm}{\centering Southern-Europe} \\
			&&&&&&\\
			\hline
			
			\multirow{2}{*}{H} & 
			\multirow{2}{*}{23} & 
			\multirow{2}{*}{Female} & 
			\multirow{2}{*}{6} & 
			\multirow{2}{*}{Multi} & 
			\multirow{2}{2cm}{\centering Hispanic} & 
			\multirow{2}{3cm}{\centering North-America \& Southeast-Asia} \\
			&&&&&&\\
			\hline
			
			\multirow{2}{*}{I} & 
			\multirow{2}{*}{27} & 
			\multirow{2}{*}{Female} & 
			\multirow{2}{*}{6} & 
			\multirow{2}{*}{Asian} & 
			\multirow{2}{2cm}{\centering Non-Hispanic} & 
			\multirow{2}{3cm}{\centering Southern-India} \\
			&&&&&&\\
			\hline
			
			\multirow{2}{*}{J} & 
			\multirow{2}{*}{25} & 
			\multirow{2}{*}{Male} & 
			\multirow{2}{*}{7} & 
			\multirow{2}{*}{Black} & 
			\multirow{2}{2cm}{\centering Non-Hispanic} & 
			\multirow{2}{3cm}{\centering Africa} \\
			&&&&&&\\
			\hline
			
			\multirow{2}{*}{K} & 
			\multirow{2}{*}{29} & 
			\multirow{2}{*}{Male} & 
			\multirow{2}{*}{7} & 
			\multirow{2}{*}{Asian} & 
			\multirow{2}{2cm}{\centering Non-Hispanic} & 
			\multirow{2}{3cm}{\centering Southern-India} \\
			&&&&&&\\
		\end{tabular}
	\end{center}
\end{table}

For all \invivo{} experiments, laser light was delivered to the clip probe (\autoref{fig:basicGeo}{(b)}) and detected from the clip probe using optical fiber bundles.
These fibers delivered light from or to an ISS Imagent V2 \gls{FD} \gls{NIRS} instrument.
The \gls{FD} \gls{NIRS} instrument used wavelengths of \SIlist{690;730;800;830}{\nano\meter}, a \SI{140.625}{\mega\hertz} modulation frequency, and a \SI{9.93}{\hertz} sample rate.
The amplitude of the \gls{FD} \gls{NIRS} data was taken as a close approximation of \gls{CW} \gls{I} in the context of this work.
\par

Eleven healthy human subjects \ie{labeled A to K} were recruited and consented according to the Tufts University \gls{IRB} protocol for this study.
The subjects' age, sex, Monk scale value,\cite{Monk_2019} race, ethnicity, and majority ancestral region are reported in \autoref{tab:subInfo}.
We report racial and ethnic information for each subject to give context on their skin tones beyond our quantification on the Monk scale.
Additionally, subjects are ordered A to K in the order of their swatch gray-scale value which corresponds to Monk scale value but with more precision (\autoref{fig:handSwatches}).
This order is used to better aid interpretation of the data in terms of skin tone.
\par

For the experimental protocol, each subject was asked to sit in a chair with their left-hand placed on a stool in-front of them, so that their hand was approximately at the height of their chest.
All of the subjects in this work reported being right-hand-dominant.
Before each experiment we ensured that the subject's hand was warm, using a space heater when necessary.
The subject placed their left-index-finger in the clip probe shown in \autoref{fig:basicGeo}{(b)}; all subjects in this study reported being right handed.
The source side of the probe corresponded to the subject's finger top \ie{the nail/knuckle side} and the detector side corresponded to the finger bottom \ie{the pad side}.
Furthermore, their fingers were placed within the clip probe such that source~1 was just behind the nail \ie{source light was not transmitted through the nail}.
Finally, \SI{3}{\minute} of \gls{NIRS} data consisting of the \gls{I} between each source and detector was collected for each subject.
These data were converted to \gls{dHbO2} and \gls{dHb} then to \gls{SpO2} for both \gls{SD} and \gls{DR} as described in \autoref{sect:meth:exp:spo2}.
\par

\section{Results}\label{sect:res} 
\subsection{Measurement path-lengths and sensitivities from Monte-Carlo simulations} 

\begin{figure}[bht]
	\begin{center}
		\includegraphics{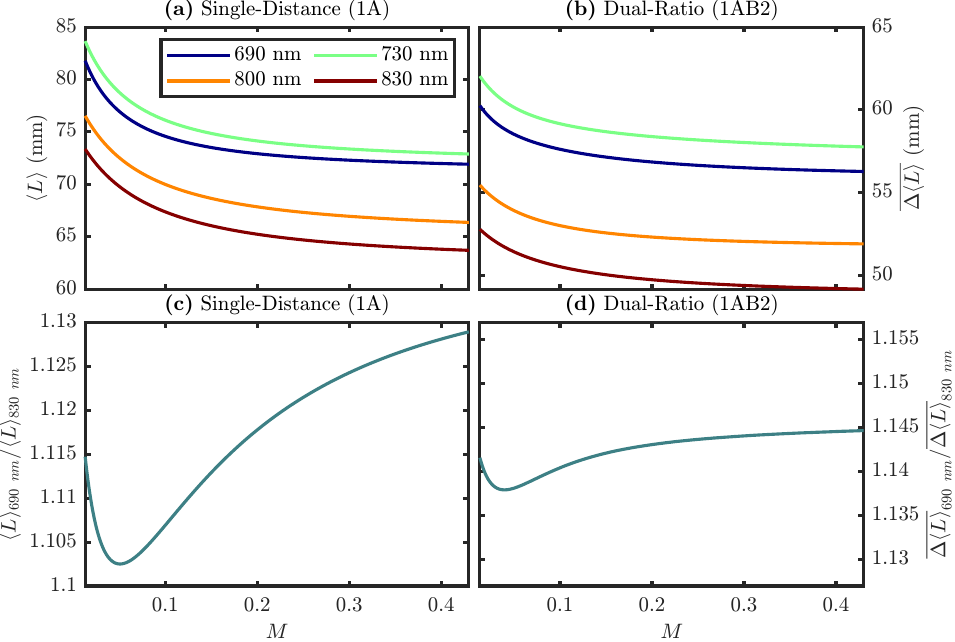}
	\end{center}
	\caption 
	{\label{fig:LanddL}
		\Acrfull{MC} derived \acrfullpl{Lpath} and \acrfullpl{dLpathAvg} for different \acrfullpl{lam} as a function of \acrfull{M}. 
		Also see \autoref{tab:LanddL}.
		{(a)} \acrshort{Lpath} for the \acrfull{SD} pair formed by source~1 and detector~A (\autoref{fig:MCgeo}).
		{(b)} \acrshort{dLpathAvg} for the \acrfull{DR} set (\autoref{fig:MCgeo}; \autoref{equ:dLpathAvg}).
		{(c)} Ratio of \acrshort{Lpath} at \SI{690}{\nano\meter} over \acrshort{Lpath} at \SI{830}{\nano\meter}.
		{(d)} Ratio of \acrshort{dLpathAvg} at \SI{690}{\nano\meter} over \acrshort{Lpath} at \SI{830}{\nano\meter}.
		\\
		Note: Subplots {(c)} \& {(d)} are on the same scale but not the same range.
	}
\end{figure}

\begin{table}[bht]
	\caption{\Acrfull{MC} derived \acrfullpl{Lpath} and \acrfullpl{dLpathAvg}; also see \autoref{fig:LanddL}}
	\label{tab:LanddL}
	\begin{center}       
		\begin{tabular}{c||c|S[table-format=2.2(1)]S[table-format=3.2(1)]S[table-format=3.2(1)]S[table-format=2.2(1)]|S[table-format=2.2(1)]} 
			{\multirow{2}{1cm}{\centering\as{M}}} & 
			{\multirow{2}{1cm}{\centering\as{lam} (\si{\nano\meter})}} & 
			{\multirow{2}{1.5cm}{\centering\as{Lpath}\textsubscript{1A} (\si{\milli\meter})}} & 
			{\multirow{2}{1.5cm}{\centering\as{Lpath}\textsubscript{1B} (\si{\milli\meter})}} & 
			{\multirow{2}{1.5cm}{\centering\as{Lpath}\textsubscript{2A} (\si{\milli\meter})}} & 
			{\multirow{2}{1.5cm}{\centering\as{Lpath}\textsubscript{2B} (\si{\milli\meter})}} & 
			{\multirow{2}{2cm}{\centering\as{dLpathAvg}\textsubscript{1AB2} (\si{\milli\meter})}} \\
			&&&&&&\\
			\hline\hline\hline
			
			\multirow{2}{*}{0.013} 
			& 690 & 81.79(2) & 142.2(2)  & 141.9(3) & 81.82(2) & 60.2(2)  \\
			& 830 & 73.38(1) & 126.03(6) & 126.3(1) & 73.37(2) & 52.78(7) \\
			
			\hline
			\multirow{2}{*}{0.430}
			& 690 & 71.92(4) & 128(1)   & 128.0(3) & 71.98(6) & 56.3(6) \\
			& 830 & 63.70(3) & 112.5(6) & 113.3(3) & 63.76(3) & 49.1(3) \\
		\end{tabular}
	\end{center}
	\small Acronyms: \Acrfull{M}, \acrfull{lam}, \acrfull{Lpath}, \acrfull{dLpathAvg} \\
	Note: Values represent the mean of three \glspl{MC} with different random seeds and the error is half the range of the three \glspl{MC}.
\end{table} 

\begin{figure}[bht]
	\begin{center}
		\includegraphics{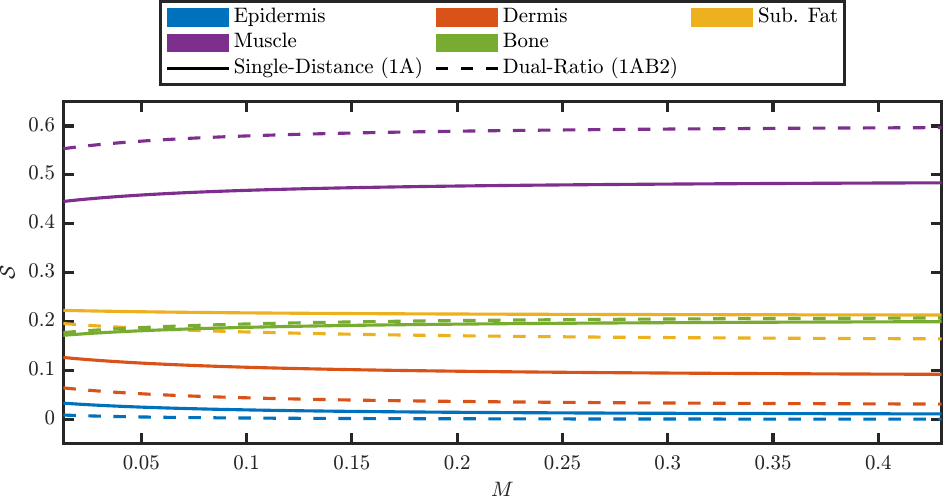}
	\end{center}
	\caption 
	{\label{fig:sensTis}
		The \acrfull{sen} at \SI{800}{\nano\meter} for the five different modeled tissues (\autoref{fig:MCgeo}) as a function of \acrfull{M}. 
		Solid lines show the \as{sen} for the \acrfull{SD} measurement type and dashed lines for \acrfull{DR} measurement type . 
	}
\end{figure}

First, we present the results from the \gls{MC} simulations.
\autoref{fig:LanddL} and \autoref{tab:LanddL} show the \gls{Lpath} from the first \gls{MC} type.
The \gls{Lpath} and the \gls{dLpathAvg} are important since they are the proportionality constants needed to convert optical data \ie{\gls{I}} to \glspl{dmua} (\autoref{equ:SDdmua} \& \autoref{equ:DRdmua}).
Focusing on \autoref{fig:LanddL}{(a)}, we see that \gls{Lpath} decreases with increasing \gls{M} in the epidermis, with a stronger decrease at lower values of \gls{M}.
This is likely due to the photons which take longer paths around the circumference of the finger having a lower probability of surviving when they encounter a more absorbing epidermis.
Next, we may examine how \gls{M} affects \gls{dLpathAvg} in \autoref{fig:LanddL}{(b)}.
Here, much like \gls{Lpath}, \gls{dLpathAvg} also decreases with increasing \gls{M}.
However, let's examine how much \gls{Lpath} or \gls{dLpathAvg} changes from the lowest to highest \gls{M} \ie{from $\as{M}=\num{0.013}$ to $\as{M}=\num{0.430}$} at \SI{830}{\nano\meter} using the values in \autoref{tab:LanddL}.
In this case \gls{Lpath} changes by \SI{-13}{\percent} while \gls{dLpathAvg} changes by \SI{-7.0}{\percent}; we remind that \gls{Lpath} is needed for \gls{SD} measurements while \gls{dLpathAvg} is needed for \gls{DR} measurements.
\par

When interpreting these values for \gls{Lpath} and \gls{dLpathAvg} in terms of their relevance toward recovering \gls{SpO2}, it is important to consider all wavelengths.
This is done in \autoref{fig:LanddL}{(c)}\&{(d)} where the ratio of \gls{Lpath} or \gls{dLpathAvg} between \SI{690}{\nano\meter} and \SI{830}{\nano\meter} is plotted.
In the case of \gls{Lpath}, this ratio would be proportional to the calibration factor applied in traditional \gls{SpO2} measurements.
This is because traditional \gls{SpO2} uses the ratio of the normalized pulsatile amplitude at red and infrared wavelengths.\cite{Nitzan_MDER14_PulseOximetry}
Here, we can also compare how much these ratios change from low to high \gls{M}.
The ratio of \glspl{Lpath} changes from \num{1.11} to \num{1.13} for $\as{M}=\num{0.013}$ to $\as{M}=\num{0.430}$, respectively; while the ratio of \glspl{dLpathAvg} changes from \num{1.142} to \num{1.145} in the same range of \gls{M}.
This is a change of \SI{1.3}{\percent} for the ratio of \gls{Lpath} and a change of \SI{0.27}{\percent} for \gls{dLpathAvg}.
This suggests that the calibration factor for \gls{DR} measurements would likely be less sensitive to \gls{M} than the factor for \gls{SD} measurements.
However, in either case these results suggest that these calibration factors change only by a few percent across \gls{M} values.
\par

\autoref{fig:sensTis} shows the \gls{sen} in different tissue regions for various values of \gls{M} and for the two measurement types, \gls{SD} or \gls{DR}.
In general both measurement types have the highest \gls{sen} to muscle and lowest \gls{sen} to epidermis. 
Comparing \gls{SD} and \gls{DR} we see that \gls{DR} is less sensitive to dynamics in superficial tissues such as dermis and epidermis but more sensitive to dynamics in deep tissues such as muscle when compared to \gls{SD}.
Additionally, looking at the dependence on \gls{M}, we see that \acrfull{sen} is lost in superficial tissues while deep tissues gain \gls{sen} at higher \gls{M} values.
However, these dependencies on \gls{M} are weak with only a change of a few percent across the full range of \gls{M} values considered.
\par

\begin{landscape}
\begin{figure}[bht]
	\begin{center}
		\includegraphics{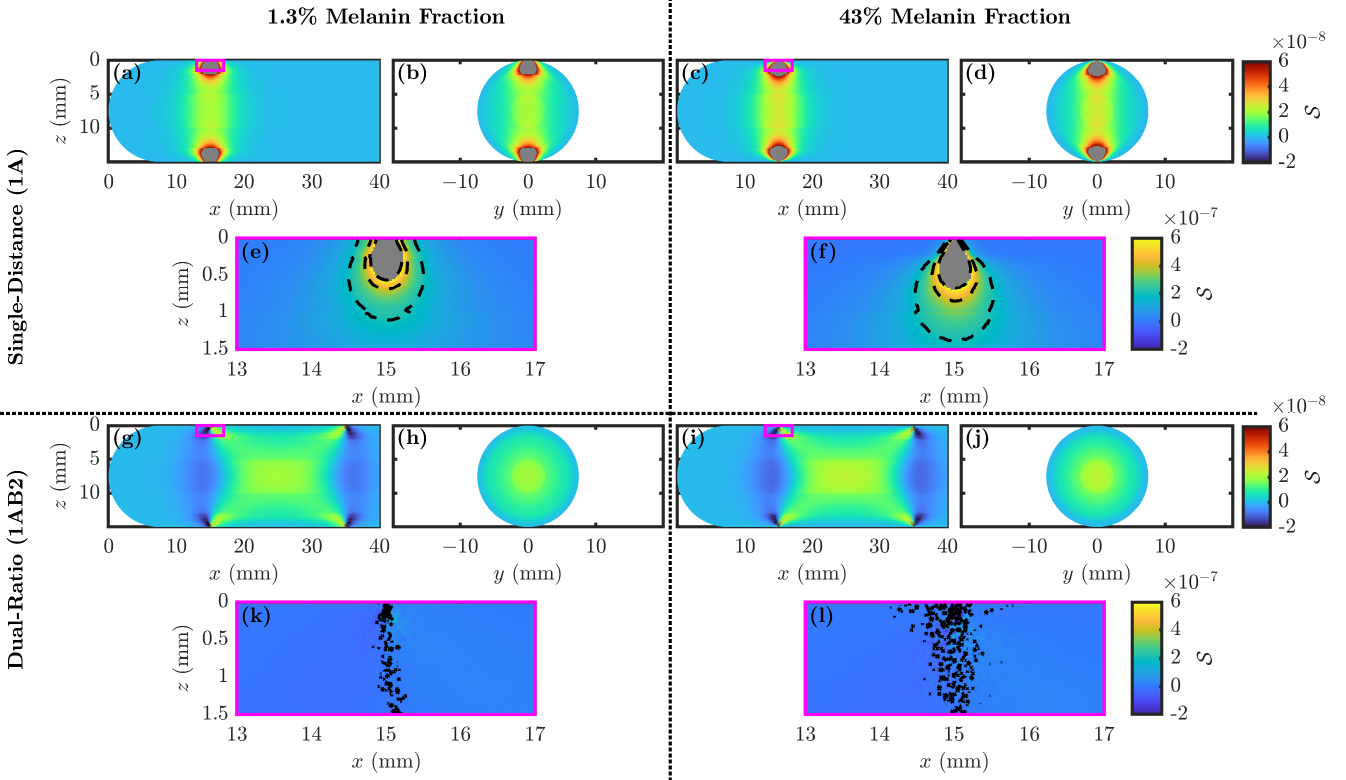}
	\end{center}
	\caption 
	{\label{fig:sensMap}
		Spatial maps of the \acrfull{sen} at \SI{800}{\nano\meter} for the \acrfull{SD} ({(a)}-{(f)}) or \acrfull{DR} ({(g)}-{(l)}) measurement types and \acrfull{M} of \num{0.013} ({(a)},{(b)},{(e)},{(g)},{(h)},\&{(k)}) or \num{0.430} ({(c)},{(d)},{(f)},{(i)},{(j)},\&{(l)}).
		Panels {(e)},{(f)},{(k)},\&{(l)} show a zoomed view of the region indicated by the magenta box in panels {(a)},{(c)},{(g)},\&{(i)}, respectively.
	}
\end{figure}
\end{landscape}

Lastly, for this section we shall look at \autoref{fig:sensMap} which contains the spatial \gls{sen} maps for the two measurement types \ie{\gls{SD} and \gls{DR}} and two \acrfull{M} values of \num{0.013} and \num{0.430}.
In general these \gls{sen} maps show a similar story to that of \autoref{fig:sensTis}, with higher values of \gls{M} resulting in a decrease in superficial \gls{sen} and a increase in deep or centralized \gls{sen}.
We can also look at the \gls{sen} very close to the optodes in \autoref{fig:sensMap}{(e)},{(f)},{(k)},\&{(l)}. 
As seen between \autoref{fig:sensMap}{(e)}\&{(f)} there is a reduction of \gls{sen} to the epidermis when a high \gls{M} is considered, which results in a deepening of the bulb of high \gls{sen} beneath the optodes for \gls{SD} measurements.
However in \autoref{fig:sensMap}{(k)}\&{(l)} we see little to no \gls{sen} to superficial tissue for the \gls{DR} measurement type regardless of \gls{M}. 
Note that \autoref{fig:sensMap}{(k)}\&{(l)} show black iso-lines which have a speckled nature, which is because many values in the zoomed map are near-zero and noise in the \gls{MC} is beginning to influence the iso-line shape.
\par

In summary, both \autoref{fig:sensTis} and \autoref{fig:sensMap} help inform where the measurements of effective \gls{dmua} come from and show that different measurement types \ie{\gls{SD} and \gls{DR}}, have different sensitivities to different tissue regions.
Therefore, if tissue hemodynamics are heterogeneous the partial volume effect governing the recovery of effective \gls{dmua} will result in different recovered hemodynamics for different measurements types. 
For this reason, in further sections, we consider hemodynamic models that contain different oscillations in different tissues to investigate the effects of this partial volume effect. 
\par

\subsection{Recovered pulsatile saturation from \invivo{} data} 

\begin{figure}[bht]
	\begin{center}
		\includegraphics{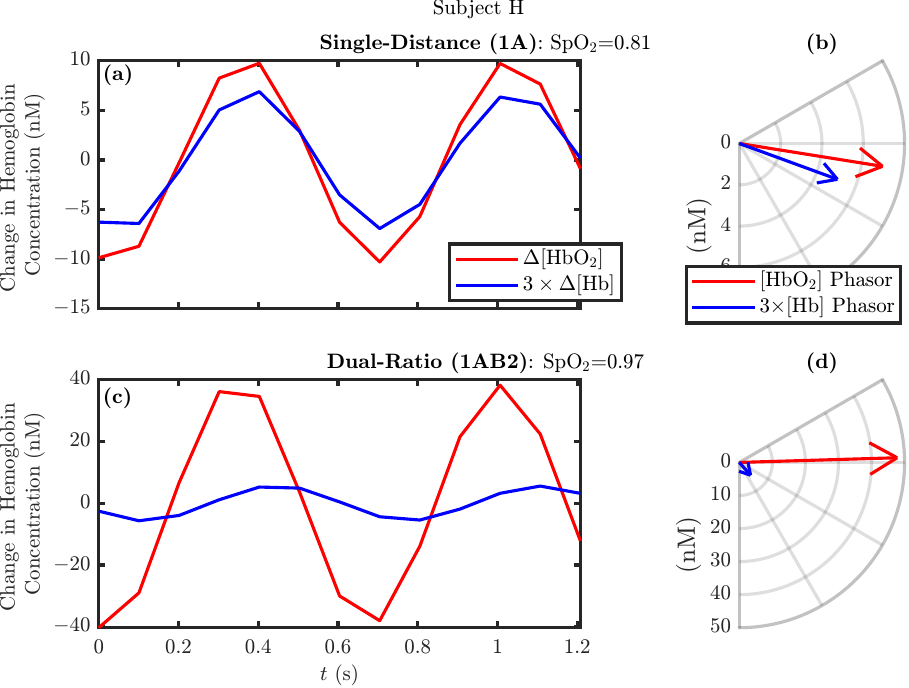}
	\end{center}
	\caption 
	{\label{fig:exPhasor}
		Example hemodynamics measured by \acrfull{SD} and \acrfull{DR} for subject~H.
		{(a)} Folding average for two periods of the band-passed temporal traces of \acrfull{dHbO2} and \acrfull{dHb} measured by \as{SD} 1A (\autoref{fig:basicGeo}).
		{(b)} Phasors for {(a)} which have the values:  $\widetilde{\as{HbO2}}=(7.0\angle\SI{-9.1}{\degree})~\si{\nano\Molar}$ and $\widetilde{\as{Hb}}=(1.7\angle\SI{-20.1}{\degree})~\si{\nano\Molar}$.
		{(c)} Same as {(a)} but measured by \as{DR} instead.
		{(d)} Phasors for {(c)} which have the values:  $\widetilde{\as{HbO2}}=(48\angle\SI{1.7}{\degree})~\si{\nano\Molar}$ and $\widetilde{\as{Hb}}=(1.7\angle\SI{-48.6}{\degree})~\si{\nano\Molar}$. \\
		Note: The assumed \acrfull{M} for this example is \num{0.013} and the phase reference is $\as{dHbO2}+\as{dHb}$ measured by \as{DR}; see \autoref{sect:meth:exp:spo2} for further details on analysis.\\
		Note: See footnote~\ref{foot:SpO2uncal} on page~\pageref{foot:SpO2uncal} regarding recovered \acrfull{SpO2}.
	}
\end{figure}

\begin{figure}[bht]
	\begin{center}
		\includegraphics{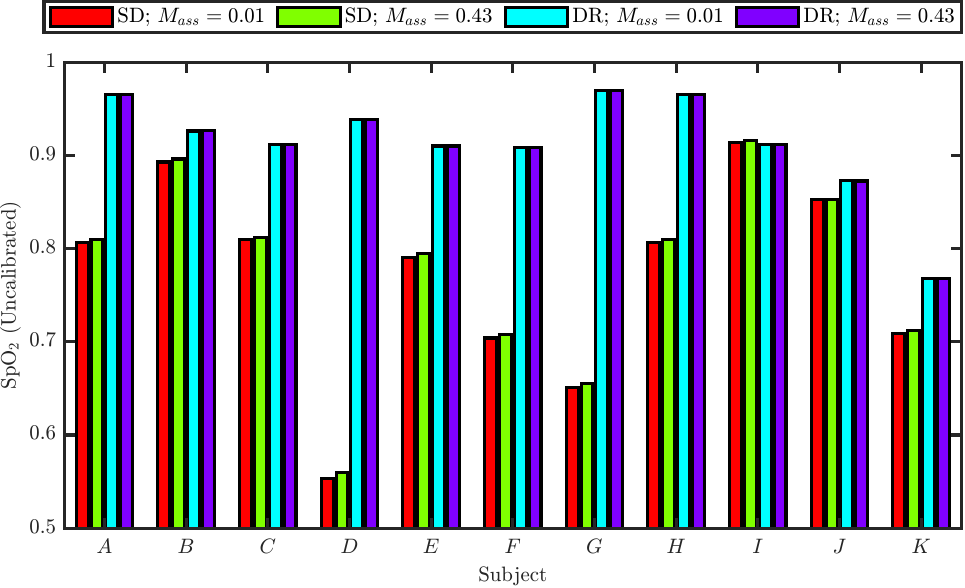}
	\end{center}
	\caption 
	{\label{fig:SpO2vsM}
		Recovered \acrfull{SpO2}\textsuperscript{\ref{foot:SpO2uncal}} \invivo{} from either \acrfull{SD} or \acrfull{DR} measurements using an assumed \acrfull{M} of \num{0.013} or \num{0.430}. Subjects are ordered from light to dark skin tones according to \autoref{sect:meth:exp:skin} \& \autoref{fig:handSwatches}.
	}
\end{figure}

Now we present the results from the \invivo{} experiments. 
\autoref{fig:exPhasor} shows example recovered hemodynamic folding average traces and phasors from subject~H.
\autoref{fig:exPhasor}{(a)}\&{(c)} contain folding average traces over two periods of the heart-frequency for the band-pass filtered \ie{with central frequency of the heart-frequency} \gls{dHbO2} and \gls{dHb} temporal traces from \gls{SD} and \gls{DR} measurement types, respectively.
While \autoref{fig:exPhasor}{(d)}\&{(d)} show the corresponding hemodynamic phasors for the oscillations in in \autoref{fig:exPhasor}{(a)}\&{(c)}, respectively.
Notice that the amplitudes of these oscillations are on the order of \si{\nano\Molar} due to the small bandwidth of the band-pass filter resulting in little remaining power in the Fourier spectrum.\footnote{For comparison the noise floor is on the order of \SI{0.1}{\nano\Molar} in the Fourier spectrum making the heart-frequency peak \gls{SNR} on the order of \num{10}.}
\par

In this work we do not consider the phase relationship of \gls{dHbO2} and \gls{dHb} when calculating \gls{SpO2}. 
However, we show the phasors including their phase relationship in \autoref{fig:exPhasor} to enable discussion of this consideration and possible future work since it is not considered in \autoref{equ:SpO2}. 
Continuing to consider the example in \autoref{fig:exPhasor}, for subject~H we see the \gls{SpO2} recovered from \gls{DR} was \SI{97}{\percent} while from \gls{SD} it was \SI{81}{\percent}.\footnote{We consider these \gls{SpO2} measurements uncalibrated since they do not represent \gls{SaO2} due to different partial volume effects, typical \gls{SpO2} techniques would effectively apply a calibration factor to these values to recover a \gls{SaO2} surrogate.\label{foot:SpO2uncal}}
A further observation is that for both \gls{SD} and \gls{DR} the \gls{Hb} phasor has a more negative phase relative to the \gls{HbO2} phasor.
This may suggest that the true tissue hemodynamics are a mixture of \gls{BV} \ie{in-phase} and \gls{BF} \ie{out-of-phase} oscillations at the heart-frequency, not solely \gls{BV} oscillations as is required for \gls{SpO2} measurements.
\par

We now move from the example data set in \autoref{fig:exPhasor} to a summary of the recovered \gls{SpO2} for all subjects in \autoref{fig:SpO2vsM}.
Here, we show the recovered \gls{SpO2} with \gls{SD} and \gls{DR} for assumed \gls{M} values of \numlist{0.013;0.430}.
In all cases, except subject~I, the recovered \gls{SpO2} from \gls{DR} was higher than the recovered \gls{SpO2} from \gls{SD}; for subject~I, the recovered \gls{SpO2} with the two measurement types was close to equal.
Furthermore, examining the dependence on the assumed value of \gls{M} we see that a higher \gls{M} increases the recovered \gls{SpO2} for \gls{SD} by a small amount; on the order of \SI{0.5}{\percent}.
While it has no noticeable effect on the recovered \gls{SpO2} with \gls{DR}.
Since \autoref{fig:SpO2vsM} shows no substantial difference between the \gls{SpO2} recovered using different assumed values of \gls{M}, the assumed values of \gls{M} have little impact for the analysis methods presented here (\autoref{sect:meth:exp:spo2}), especially in the case of \gls{DR}.
\par

We may also examine \autoref{fig:SpO2vsM} in terms of skin tone dependence.
As a reminder, we have ordered the subjects A to K in order from lightest to darkest skin tone according to the methods in \autoref{sect:meth:exp:skin} and \autoref{fig:handSwatches}.
The \gls{SD} measurements appear to have little to no dependence on skin tone, though the values of recovered \gls{SpO2} vary greatly with a minimum of \SI{55}{\percent} for for subject~D to a maximum of \SI{91}{\percent} for subject~I.
By contrast, the variation of recovered \gls{SpO2} values for \gls{DR} across subjects is much less, with a minimum of \SI{77}{\percent} for subject~k and a maximum of \SI{97}{\percent} for subject~G.
We also point out that \autoref{fig:SpO2vsM} may show a dependence on skin tone for \gls{DR} measurements if we look at the recovered \gls{SpO2} values from subject~H to subject~K.
This may suggest an affect that results in a lower recovered \gls{SpO2} for darker skin when using \gls{DR}.
This would represent a negative bias which is at odds with results for conventional pulse-oximetry in the literature.\cite{SjodingMichaelW._N.Engl.J.Med.20_RacialBias, Moradi_Des.Qual.Biomed.Technol.XVII24_ModelingLighttissue}
These lower \glspl{SpO2} recovered by \gls{DR}, particularly for subjects~J~and~K, aren't corrected for by assuming relevant \gls{M} \ie{there is no noticeable difference between assuming \gls{M} of \num{0.013} or \num{0.430} for any subject}.
We caution that this dependence observed in subjects H to K is tenuous due to the small number of subjects and the observed variations in \gls{SpO2} across all subjects.
We only point out this dependence as a possible point of discussion that will be further investigated in future work, which will also consider the phase differences between \gls{dHbO2} and \gls{dHb} (\autoref{fig:exPhasor}).
\par

\subsection{Recovered saturation from Monte-Carlo simulations informed by \invivo{} results} 

\begin{figure}[bht]
	\begin{center}
		\includegraphics{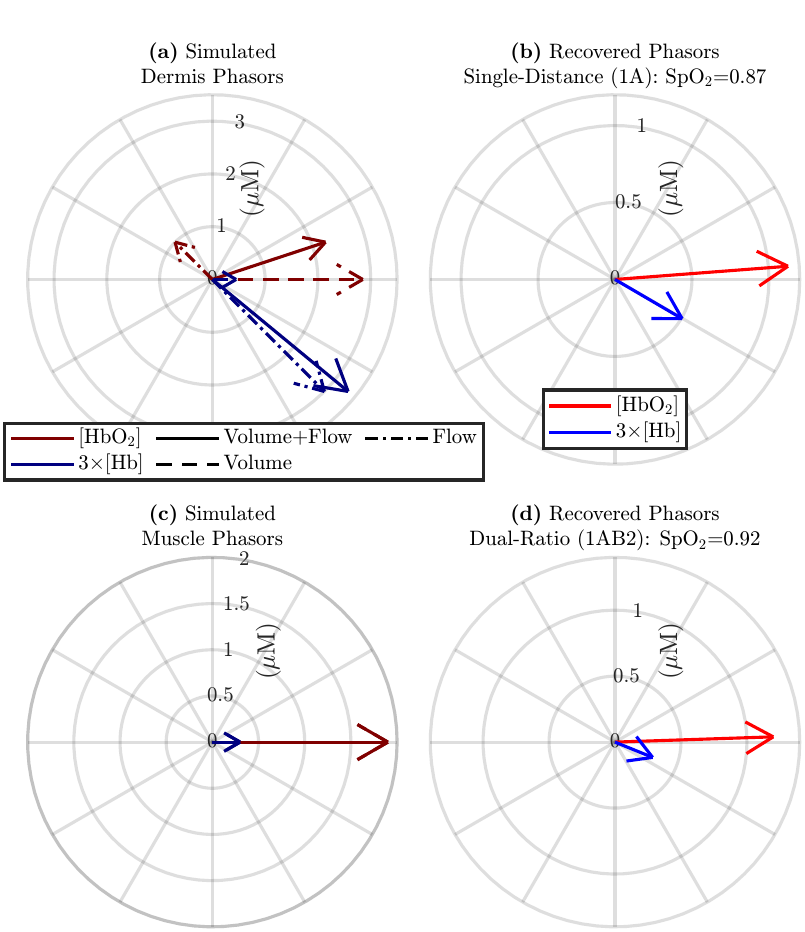}
	\end{center}
	\caption 
	{\label{fig:simPhasors}
		Simulation of tissue \acrfull{HbO2} and \acrfull{Hb} phasors leading to recovered phasors and \acrfull{SpO2} measured by \acrfull{SD} or \acrfull{DR}. Hemodynamics are only in the dermis (\acrfull{BV} and \acrfull{BF}) and muscle (only \as{BV}). \as{BV} oscillations have a saturation of \SI{95}{\percent}. \Acrfull{M} modeled as \num{0.013}.
		{(a)} Dermis phasors: 
		$(2.26\angle\SI{18.3}{\degree})~\si{\micro\Molar}$ for \gls{HbO2} and  $(1.11\angle\SI{-39.5}{\degree})~\si{\micro\Molar}$ for \gls{Hb}.
		{(b)} \as{SD} recovered phasors: 
		$(1.13\angle\SI{4.4}{\degree})~\si{\micro\Molar}$ for \gls{HbO2} and  $(0.17\angle\SI{-30.3}{\degree})~\si{\micro\Molar}$ for \gls{Hb}.
		{(c)} Muscle phasors: 
		$(1.90\angle\SI{0}{\degree})~\si{\micro\Molar}$ for \gls{HbO2} and  $(0.10\angle\SI{0}{\degree})~\si{\micro\Molar}$ for \gls{Hb}.
		{(d)} \as{DR} recovered phasors: 
		$(1.20\angle\SI{1.9}{\degree})~\si{\micro\Molar}$ for \gls{HbO2} and  $(0.10\angle\SI{-22.1}{\degree})~\si{\micro\Molar}$ for \gls{Hb}.
	}
\end{figure}

\begin{figure}[bht]
	\begin{center}
		\includegraphics{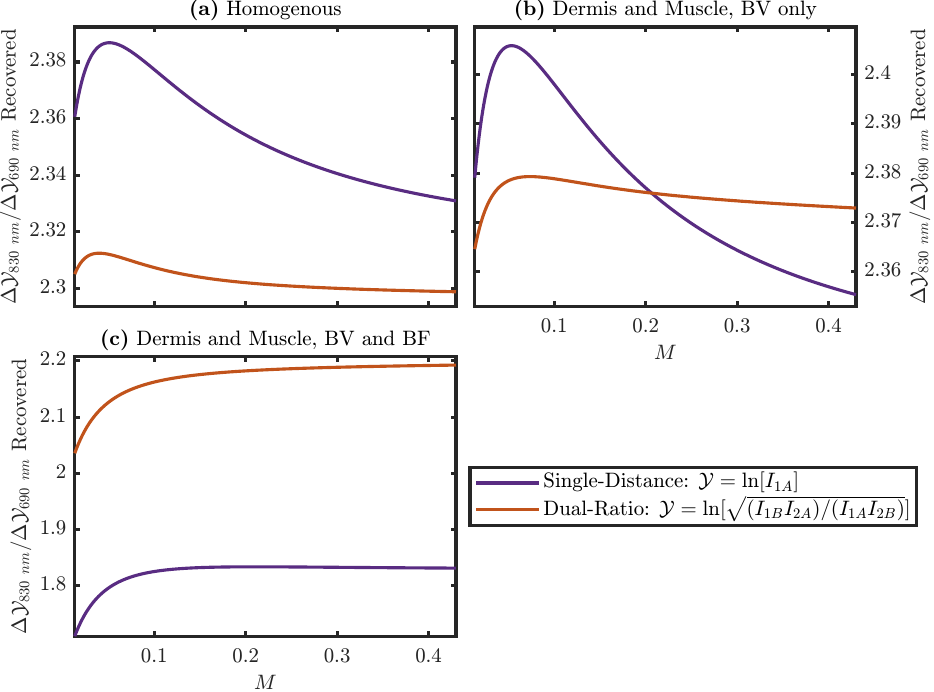}
	\end{center}
	\caption 
	{\label{fig:simAssM_dIdI}
		Simulated ratios of the changes in \acrfull{Y} at \SIlist{830;690}{\nano\meter} as a function of modeled \gls{M}. 
		Changes in \as{Y} are the numerators of \autoref{equ:SDdmua} or \autoref{equ:DRdmua} which are also written in the legend for either \acrfull{SD} or \acrfull{DR}.
		{(a)} Simulation with homogeneous \acrfull{BV} oscillations in the whole tissue with phasor values of $(0.95\angle\SI{0}{\degree})~\si{\micro\Molar}$ for \acrfull{HbO2} and $(0.05\angle\SI{0}{\degree})~\si{\micro\Molar}$ for \acrfull{Hb}.
		{(b)} Simulation with \as{BV} oscillations only in the dermis and muscle again with phasor values of $(0.95\angle\SI{0}{\degree})~\si{\micro\Molar}$ for \as{HbO2} and $(0.05\angle\SI{0}{\degree})~\si{\micro\Molar}$ for \as{Hb}.
		{(c)} Simulation with \acrfull{BF} and \as{BV} oscillations in the dermis and only \as{BV} in the muscle. 
		This is the same simulation as \autoref{fig:simPhasors}, and the simulated phasor values may be found there.
	}
\end{figure}

\begin{figure}[bht]
	\begin{center}
		\includegraphics{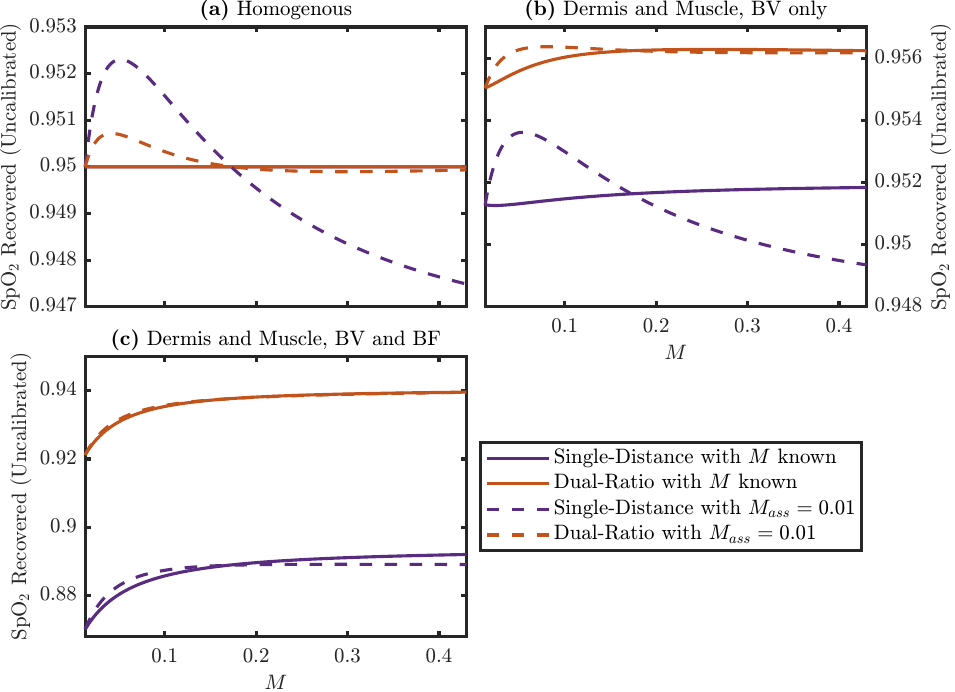}
	\end{center}
	\caption 
	{\label{fig:simAssM}
		Simulated recovered \acrfull{SpO2} for \acrfull{SD} or \acrfull{DR}. Solid lines are the case where the true \acrfull{M} is known and used to recover \gls{SpO2}, while dashed lines assume a value for \as{M} of \num{0.013}.
		For all simulations the true \acrfull{SaO2} from \acrfull{BV} oscillations is \SI{95}{\percent}.
		{(a)} Simulation with homogeneous \as{BV} oscillations in the whole tissue with phasor values of $(0.95\angle\SI{0}{\degree})~\si{\micro\Molar}$ for \acrfull{HbO2} and $(0.05\angle\SI{0}{\degree})~\si{\micro\Molar}$ for \acrfull{Hb}.
		{(b)} Simulation with \as{BV} oscillations only in the dermis and muscle again with phasor values of $(0.95\angle\SI{0}{\degree})~\si{\micro\Molar}$ for \as{HbO2} and $(0.05\angle\SI{0}{\degree})~\si{\micro\Molar}$ for \as{Hb}.
		{(c)} Simulation with \acrfull{BF} and \as{BV} oscillations in the dermis and only \as{BV} in the muscle. 
		This is the same simulation as \autoref{fig:simPhasors}, and the simulated phasor values may be found there.
	}
\end{figure}

As described in \autoref{sect:meth:MC:simSpO2}, we may simulate either \gls{SD} or \gls{DR} measurements for different tissue hemodynamics and \gls{M} values based on the \gls{MC} model.
Furthermore, we can simulate measurements for given hemodynamics and a value for \gls{M} but analyze the data with a different assumed value of \gls{M} to investigate the effect of such assumption.
Of course, these simulations are with the caveat of being within the context of the \gls{MC} finger model in this work.
\par

One consistent result of the \invivo{} measurements reported in \autoref{fig:SpO2vsM} was that \gls{SpO2} recovered by \gls{DR} was greater than the one recovered by \gls{SD}. 
This may be explained by different hemodynamics in different tissues where \gls{SD} and \gls{DR} have different \acrfull{sen} making the partial volume effect come into play (\autoref{fig:sensTis} and \autoref{fig:sensMap}).
We report a simulation of hemodynamic phasors at the heart-frequency that recreate this consistent result.
To that end, we may model hemodynamics as combinations of \gls{BV} and \gls{BF}, with \gls{HbO2} and \gls{Hb} phasors being in-phase if driven by \gls{BV} oscillations and out-of-phase if driven by \gls{BF} oscillations.
Further, the relative amplitudes of the arterial \gls{BV} components of the \gls{HbO2} and \gls{Hb} phasors is \gls{SaO2}.
For \gls{BF} components, the amplitudes of \gls{HbO2} and \gls{Hb} phasors are equal.
Considering this, we modeled hemodynamics as a combination of \gls{BF} and \gls{BV} with an \gls{SaO2} of \SI{95}{\percent}. 
Oscillations were assumed to be in the dermis and muscle only, with only the dermis having a \gls{BF} contribution.
In the dermis the \gls{BF} components were $(1\angle\SI{135}{\degree})~\si{\micro\Molar}$ for the \gls{HbO2} phasor and $(1\angle\SI{-45}{\degree})~\si{\micro\Molar}$ for the \gls{Hb} phasor, while \gls{BV} components were $(2.85\angle\SI{0}{\degree})~\si{\micro\Molar}$ for the \gls{HbO2} phasor and $(0.15\angle\SI{0}{\degree})~\si{\micro\Molar}$ for the \gls{Hb} phasor (\autoref{fig:simPhasors}{(a)}).
Meanwhile, for the muscle, the \gls{BV} components were $(1.90\angle\SI{0}{\degree})~\si{\micro\Molar}$ for the \gls{HbO2} phasor and  $(0.10\angle\SI{0}{\degree})~\si{\micro\Molar}$ for the \gls{Hb} phasor (\autoref{fig:simPhasors}{(c)}).
Given this simulation and the model, the recovered phasors are shown in \autoref{fig:simPhasors}{(b)}\&{(d)} which recreate the greater \gls{SpO2} recovered by \gls{DR} which was observed \invivo{}.
We also note that the simulation in \autoref{fig:simPhasors} qualitatively recreates the phase relationships between \gls{HbO2} and \gls{Hb} phasors shown in the example dataset in \autoref{fig:exPhasor}, though we do not put much weight on this agreement since it is with one example subject.
\par

We can extend these simulations by investigating the assumed value of \gls{M} for \gls{SpO2}.
For this, we consider three different hemodynamic simulations \ie{two simulations in addition to the one already described}.
One hemodynamic simulation is the one in \autoref{fig:simPhasors} which recreated some aspects of the \invivo{} results and considers a simulation of \gls{BV} and \gls{BF} in the dermis and only \gls{BV} in the muscle.
The first additional simulation assumed a homogeneous \gls{BV} oscillation and \gls{SaO2} of \SI{0.95}{\percent} in all tissues, with no \gls{BF} oscillations anywhere.
That is an \gls{HbO2} phasor of $(0.95\angle\SI{0}{\degree})~\si{\micro\Molar}$ and an \gls{Hb} phasor of $(0.05\angle\SI{0}{\degree})~\si{\micro\Molar}$ in all tissues.
This homogeneous case will only show the effect of different assumed values of \gls{M}.
The second additional simulation is similar to the first, with only \gls{BV} oscillations, but now only in the dermis and muscle and no hemodynamics at the heart-frequency in the other tissues.
This case also considered an \gls{HbO2} phasor of $(0.95\angle\SI{0}{\degree})~\si{\micro\Molar}$ and an \gls{Hb} phasor of $(0.05\angle\SI{0}{\degree})~\si{\micro\Molar}$, but now only in the dermis and muscle.
\par

\autoref{fig:simAssM_dIdI} shows simulated recovered data for the three hemodynamic simulations presented in a similar way to the data collected for traditional \gls{SpO2} methods, as ratios of pulsatile components of \gls{dY} at two wavelengths. 
In \autoref{fig:simAssM_dIdI} the ratio of changes in \gls{dY} at \SIlist{830;690}{\nano\meter} are plotted versus \gls{M}.
For \gls{SD} and \gls{DR} changes in \gls{dY} is given by the numerators of the right-hand-side of \autoref{equ:SDdmua} and \autoref{equ:DRdmua}, respectively.
The three panels of \autoref{fig:simAssM_dIdI} represent the three simulations.
One will note that, the recovered data for both \gls{SD} and \gls{DR} does depend on \gls{M}, but the dependence on \gls{M} is stronger for \gls{SD} in the two simulations which assume no \gls{BF} oscillations \ie{the first and second}.
\par

Finally, we come to \autoref{fig:simAssM} which conveys the same information as \autoref{fig:simAssM_dIdI} but with the added information of what the recovered \gls{SpO2} would be for different cases of assumed \gls{M}. \autoref{fig:simAssM} considers all four wavelengths but \autoref{fig:simAssM_dIdI} only considers \SIlist{690;830}{\nano\meter}.
The three simulations described above are shown in \autoref{fig:simAssM}, and the dashed lines represent simulated data analyzed with an assumed \gls{M} of \num{0.013}.
In all three simulations the true \gls{SaO2} \ie{the saturation of the \gls{BV} oscillation} is \SI{95}{\percent}.
The true \gls{M} in the simulation \ie{the on used to generate the forward data} corresponds to the value on the $x$-axis and solid lines are recovered \gls{SpO2} using this true \gls{M} value.
First, let's look at \autoref{fig:simAssM}{(a)} which shows the simulation with homogeneous \gls{BV} oscillations in all the tissue regions.
The solid lines in \autoref{fig:simAssM}{(a)} match the simulated \gls{SaO2} and recover a \gls{SpO2} of \SI{95}{\percent} which verifies the validity of the methods in this work.
Note that the solid lines in \autoref{fig:simAssM}{(a)} are coincident.
Further the deviation of the dashed line from the solid line shows the effect of assuming an incorrect value of \gls{M} without any partial-volume effects confounding the simulation.
From this we see that the recovered \gls{SpO2} from \gls{SD} deviates from the true value by less than \SI{0.25}{\percent}, while the recovered \gls{SpO2} from \gls{DR} deviates by less than \SI{0.1}{\percent}.
Moving to the second simulation in \autoref{fig:simAssM}{(b)}, which only considers \gls{BV} oscillations that are in the dermis and muscle, we see a similar story to \autoref{fig:simAssM}{(a)} but with a systematic shift for the \gls{SD} or \gls{DR} recovered \gls{SpO2}.
In both cases \gls{SpO2} overestimates the simulated \gls{SaO2}, with \gls{DR} overestimating it more, with the difference at about \SI{0.6}{\percent}.
These shifts of the solid lines in \autoref{fig:simAssM}{(b)} results from partial volume effects.
Finally, we move to the last simulation in \autoref{fig:simAssM}{(c)}, which is the same as \autoref{fig:simPhasors}.
In this case, there is a large difference between \gls{SpO2} recovered by \gls{SD} and \gls{DR}.
This result matches the \invivo{} data in that the value for \gls{DR} is greater regardless of \gls{M}.
Additionally, the recovered \gls{SpO2} varies by about \SI{2}{\percent} across \gls{M} for both measurement types and this is not corrected for by knowing the true \gls{M} when analyzing the data \ie{the solid curve varies more than the difference between the solid and dashed curves}.
Since the recovered \gls{SpO2} in \autoref{fig:simAssM}{(c)} increases with increasing \gls{M}, this is at odds with the decrease observed between subjects H and K as shown in \autoref{fig:SpO2vsM}, further suggesting that this observation from \autoref{fig:SpO2vsM} may not be significant.
\par

Wrapping up these simulations, we can see that the partial volume effect and heterogeneous tissue hemodynamic phasors affect \gls{SpO2} values recovered by both \gls{SD} and \gls{DR} to a greater extent than different values of \gls{M}, whether or not the correct value of \gls{M} is known.
Furthermore, in the simulation which roughly recreated the \invivo{} results (\autoref{fig:simAssM}{(c)}) \gls{DR} recovered a \gls{SpO2} value closer to the modeled \gls{SaO2}.
This is consistent with the higher values of \gls{SpO2} found by \gls{DR} in the \invivo{} data presented in \autoref{fig:SpO2vsM}.
But these trends in \autoref{fig:simAssM} show changes in recovered \gls{SpO2} that are much less than the observed differences in recovered \gls{SpO2} between different subjects in \autoref{fig:SpO2vsM}.
These greater differences \invivo{} suggest that the simulations may not be capturing the full picture.
We conclude this section by reminding the readers that these results are all within the context of the \gls{MC} model we used in this work, which assumes a particular anatomy, optical properties, and model of skin tone based on \gls{M}.
\par

\section{Discussion}\label{sect:DisCon} 

In this work, we presented a \gls{MC} model to simulate optical measurements on the human finger and experimental results obtained \invivo{} using the same optical measurement geometry (\autoref{fig:basicGeo}).
We designed the finger \gls{MC} model based on a simplified geometry of tissue types in a human finger and assumed optical properties based on previous literature.\cite{Jacques_PMB13_OpticalProperties, Jacques_PMB13_ErratumOptical} 
Then we utilized this model to analyze the \invivo{} data and recover \gls{SpO2} for two different measurement types \ie{\gls{SD} and \gls{DR}}.
The results of the \invivo{} experiment were then used to inform models of hemodynamic oscillations and enable a discussion of the differences between measurement types, assumed \gls{M}, and heterogeneous tissue hemodynamics.
\par

The \gls{MC} simulations culminated in the results of \autoref{fig:simAssM} for three different tissue hemodynamic models.
These models considered hemodynamic oscillations at the heart-frequency in different tissue regions with different phase and amplitude relationships.
One hemodynamic model was created to roughly match some consistent results obtained in the \invivo{} measurements. 
In contrast, two other hemodynamic models were used to investigate the interplay between tissue hemodynamic heterogeneity and assumed values of \gls{M}.
Overall, the results showed that the assumed value for \gls{M}, whether assumed incorrectly or correctly, affected the recovered \gls{SpO2} on the order of \SI{1}{\percent}.
However, the modeled heterogeneity of tissue hemodynamics significantly affected the recovered \gls{SpO2}; this effect was greater than the effect from assumed \gls{M}.
The different results obtained with different measurement types are due to different partial volume effects and show evidence of spatially heterogeneous tissue hemodynamics.
\par

Expounding upon these ideas regarding assumed values of \gls{M} or spatially varying tissue hemodynamics, we can focus on \autoref{fig:LanddL}, \autoref{fig:simAssM_dIdI}, and \autoref{fig:simAssM}. 
\autoref{fig:LanddL} represents the proportionality constants modeled by the \gls{MC} simulations, which govern how the change in the measured optical data are converted to the \gls{dmua}, which would later be converted to the \gls{dHbO2} and \gls{dHb}, and then finally \gls{SpO2}. 
The simulations reported in \autoref{fig:simAssM_dIdI} show that the measured optical data are affected by different hemodynamic situations and \gls{M} values. 
Finally, \autoref{fig:simAssM} shows the simulated recovered \gls{SpO2} for the same hemodynamic situations as in \autoref{fig:simAssM_dIdI}.
The lines in \autoref{fig:simAssM} effectively result from a multiplication of the values in \autoref{fig:LanddL}{(c)}\&{(d)} and \autoref{fig:simAssM_dIdI}.\footnote{This statement is not formally true since \autoref{fig:LanddL}{(c)}\&{(d)} and \autoref{fig:simAssM_dIdI} consider two wavelengths but \autoref{fig:simAssM} uses all four wavelengths in this work.}
This relationship, between the values of \autoref{fig:LanddL}, \autoref{fig:simAssM_dIdI}, and \autoref{fig:simAssM}, shows the interplay between the measured optical data and assumed calibration constants needed to obtain \gls{dmua}, which is related to \gls{SpO2} through known extinction coefficients.\cite{Prahl_98_TabulatedMolar} 
Effectively, the curves in \autoref{fig:LanddL}{(c)}\&{(d)} are the calibration constants that would need to be found to convert optical data to a measurement proportional to \gls{SpO2} and therefore their variation over \gls{M} shows how these calibration constants may need to change with skin tone.
Meanwhile, \autoref{fig:simAssM_dIdI} and \autoref{fig:simAssM} add the consideration of heterogeneous tissue hemodynamic oscillations and how the recovered measurements would change with them.
\par

Looking at all of these together, we see that the recovered \gls{SpO2} was more influenced by hemodynamic heterogeneity than assumed \gls{M}.
Since different people will likely have different hemodynamic heterogeneity we will expect that the recovered \gls{SpO2} will vary more across subjects than across assumed \gls{M} values within a subject.
This is in-fact observed in \autoref{fig:SpO2vsM} where variation across subject is much greater than the differences obtained by assuming different \gls{M} values.
Furthermore, the variation across subjects is greater for \gls{SD} compared to \gls{DR} suggesting that \gls{DR}, at least in-part, compensates for the subject differences.
We can put these results in the context of Reference~\citenum{SjodingMichaelW._N.Engl.J.Med.20_RacialBias} which showed a bias in the \gls{SpO2} measured on Black versus White patients. In Reference~\citenum{SjodingMichaelW._N.Engl.J.Med.20_RacialBias} the variation across patients \ie{for a given value of \gls{SaO2}} was greater than the observed skin tone bias.
This skin tone bias, in Reference~\citenum{SjodingMichaelW._N.Engl.J.Med.20_RacialBias}, is evident when considering the large patient population which, when averaged, likely suppressed the hemodynamic or physiological differences between the patients making the effect of skin tone appear.
Our results are in-line with this interpretation of the results in Reference~\citenum{SjodingMichaelW._N.Engl.J.Med.20_RacialBias} given that we expect more variation in \gls{SpO2} from hemodynamic differences than skin tone \ie{modeled as \gls{M}} differences.
\par

One criticism of traditional \gls{SpO2} measurements is that the calibration factor is assumed to be the same for all skin tones and thus \gls{M}.\cite{Setchfield_JBO24_EffectSkin, Chatterjee_Sensors19_MonteCarlo}
For the results obtained from the models, simulations, and analysis methods reported in this article, we see that this assumption creates only small inaccuracies in \gls{SpO2}, on the order of \SI{1}{\percent}. 
Instead, the results in this work show that the larger confound on recovered \gls{SpO2} measurements is the heterogeneity of tissue hemodynamics. 
\par

Beyond the discussion of \gls{M} and tissue heterogeneity, we also explore the difference between measurement types, \gls{SD} and \gls{DR}.
One consistent result that we see from the \invivo{} data was a higher \gls{SpO2} recovered by \gls{DR} compared to \gls{SD}.
More importantly, the \gls{SpO2} recovered by \gls{DR} was more consistent across these healthy human subjects, which is consistent with these subjects having approximately the same nominal \gls{SaO2}. 
This result was recreated in a simulation that modeled pulsatile \gls{BV} in the dermis and muscle tissue, with pulsatile \gls{BF} in only occurring in the dermis.
However, the hemodynamic model and relationship of tissue sensitivities are complex, so we do not claim that this simulation, which recreated the \invivo{} data relationships, is necessarily representative of the actual tissue dynamics.
While this case may be possible, other situations may also reproduce our \invivo{} results.
\par 

\section{Conclusion}\label{sect:Con} 

In conclusion, our results indicate that optical measurements of \gls{SpO2} may be more dependent on heterogeneity in tissue hemodynamics than on the assumed value of \gls{M}. 
Furthermore, we found that \gls{DR} measurements recovered greater and more consistent \gls{SpO2} values than \gls{SD} measurements across various subjects.
These results will further enable discussions about optimal methods to recover \gls{SpO2} with minimal impact from skin tone.
Further, these results show promise of the \gls{DR} method for \gls{SpO2} measurements on the human finger.
Future directions include considering the amplitude and phase relationships of pulsatile hemodynamics of \gls{dHbO2} and \gls{dHb} when measuring \gls{SpO2}, finding ways to better assess and account for different \glspl{M} and skin tones, investigating pulsatile hemodynamic heterogeneity in the finger, considering the number and values of wavelengths used, using of novel measurement methods such as \gls{DR}, and considering time-resolved optical changes in the time or frequency domain.
We hope that this work may represent a jumping off point for future work on these considerations.
\par 


\subsection*{Disclosures}
The authors disclose no conflicts of interest.

\subsection*{Data, Materials, and Code Availability} 
Applicable supporting code and data are available from the authors upon reasonable request. 

\subsection*{Acknowledgments}
This work is supported by \gls{NIH} award R01-EB029414.
G.B. would also like to acknowledge support from \gls{NIH} award K12-GM133314.
The content is solely the authors' responsibility and does not necessarily represent the official views of the awarding institutions.
\par

We would also like to acknowledge helpful correspondence with Qianqian Fang regarding Monte-Carlo~eXtreme.

\bibliography{GBlib240426}   

\begin{thebibliography}{10}

\bibitem{Tremper_Chest89_PulseOximetry}
K.~K. Tremper, ``Pulse oximetry,'' {\em Chest} {\bf 95}(4), 713--715  (1989).

\bibitem{Severinghaus_Anesth.Analg.07_TakuoAoyagi}
J.~W. Severinghaus, ``Takuo {{Aoyagi}}: {{Discovery}} of {{Pulse Oximetry}},''
  {\em Anesthesia \& Analgesia} {\bf 105}, S1  (2007).
\newblock
  \url{https://journals.lww.com/anesthesia-analgesia/fulltext/2007/12001/takuo_aoyagi__discovery_of_pulse_oximetry.1.aspx}.

\bibitem{Nitzan_MDER14_PulseOximetry}
M.~Nitzan, A.~Romem, and R.~Koppel, ``Pulse oximetry: Fundamentals and
  technology update,'' {\em Medical Devices: Evidence and Research} {\bf 7},
  231--239  (2014).
\newblock
  \url{https://www.dovepress.com/pulse-oximetry-fundamentals-and-technology-update-peer-reviewed-fulltext-article-MDER}.

\bibitem{Chan_RespiratoryMedicine13_PulseOximetry}
E.~D. Chan, M.~M. Chan, and M.~M. Chan, ``Pulse oximetry: {{Understanding}} its
  basic principles facilitates appreciation of its limitations,'' {\em
  Respiratory Medicine} {\bf 107}, 789--799  (2013).
\newblock
  \url{https://www.sciencedirect.com/science/article/pii/S095461111300053X}.

\bibitem{Leppanen_Advancesinthediagnosisandtreatmentofsleepapnea:Fillingthegapbetweenphysiciansandengineers22_PulseOximetry}
T.~Lepp{\"a}nen, S.~Kainulainen, H.~Korkalainen, {\em et~al.}, ``Pulse
  oximetry: {{The}} working principle, signal formation, and applications,'' in
  {\em Advances in the Diagnosis and Treatment of Sleep Apnea: {{Filling}} the
  Gap between Physicians and Engineers},  205--218, Springer  (2022).

\bibitem{Millikan_RSI42_OximeterInstrument}
G.~A. Millikan, ``The {{Oximeter}}, an {{Instrument}} for {{Measuring
  Continuously}} the {{Oxygen Saturation}} of {{Arterial Blood}} in {{Man}},''
  {\em Review of Scientific Instruments} {\bf 13}, 434--444  (1942).
\newblock \url{https://doi.org/10.1063/1.1769941}.

\bibitem{Setchfield_JBO24_EffectSkin}
K.~Setchfield, A.~Gorman, A.~H. R.~W. Simpson, {\em et~al.}, ``Effect of skin
  color on optical properties and the implications for medical optical
  technologies: A review,'' {\em Journal of Biomedical Optics} {\bf 29}, 010901
   (2024).
\newblock
  \url{https://www.spiedigitallibrary.org/journals/journal-of-biomedical-optics/volume-29/issue-1/010901/Effect-of-skin-color-on-optical-properties-and-the-implications/10.1117/1.JBO.29.1.010901.full}.

\bibitem{Al-Halawani_Physiol.Meas.23_ReviewEffect}
R.~{Al-Halawani}, P.~H. Charlton, M.~Qassem, {\em et~al.}, ``A review of the
  effect of skin pigmentation on pulse oximeter accuracy,'' {\em Physiological
  Measurement}   (2023).

\bibitem{Shi_BMCMed.22_AccuracyPulse}
C.~Shi, M.~Goodall, J.~Dumville, {\em et~al.}, ``The accuracy of pulse oximetry
  in measuring oxygen saturation by levels of skin pigmentation: A systematic
  review and meta-analysis,'' {\em BMC medicine} {\bf 20}(1), 267  (2022).

\bibitem{Cabanas_Sensors22_SkinPigmentation}
A.~M. Cabanas, M.~{Fuentes-Guajardo}, K.~Latorre, {\em et~al.}, ``Skin
  pigmentation influence on pulse oximetry accuracy: A systematic review and
  bibliometric analysis,'' {\em Sensors} {\bf 22}(9), 3402  (2022).

\bibitem{Bierman_Br.J.Anaesth.24_MelaninBias}
A.~Bierman, K.~Benner, and M.~S. Rea, ``Melanin bias in pulse oximetry
  explained by light source spectral bandwidth,'' {\em British Journal of
  Anaesthesia}   (2024).

\bibitem{Martin_Br.J.Anaesth.24_EffectSkin}
D.~Martin, C.~Johns, L.~Sorrell, {\em et~al.}, ``Effect of skin tone on the
  accuracy of the estimation of arterial oxygen saturation by pulse oximetry: A
  systematic review,'' {\em British Journal of Anaesthesia}   (2024).

\bibitem{Mantri_Biomed.Opt.Express22_ImpactSkin}
Y.~Mantri and J.~V. Jokerst, ``Impact of skin tone on photoacoustic oximetry
  and tools to minimize bias,'' {\em Biomedical Optics Express} {\bf 13}(2),
  875--887  (2022).

\bibitem{Moradi_Des.Qual.Biomed.Technol.XVII24_ModelingLighttissue}
M.~Moradi, S.~Vasudevan, A.~Bhusal, {\em et~al.}, ``Modeling light-tissue
  interactions in pulse oximetry: Effect of device design and skin
  pigmentation,'' in {\em Design and {{Quality}} for {{Biomedical Technologies
  XVII}}},   {\bf 12833}, 1283302, SPIE  (2024).
\newblock
  \url{https://www.spiedigitallibrary.org/conference-proceedings-of-spie/12833/1283302/Modeling-light-tissue-interactions-in-pulse-oximetry--effect-of/10.1117/12.3004189.full}.

\bibitem{SjodingMichaelW._N.Engl.J.Med.20_RacialBias}
{Sjoding Michael W.}, {Dickson Robert P.}, {Iwashyna Theodore J.}, {\em
  et~al.}, ``Racial {{Bias}} in {{Pulse Oximetry Measurement}},'' {\em New
  England Journal of Medicine} {\bf 383}, 2477--2478  (2020).
\newblock \url{https://www.nejm.org/doi/full/10.1056/NEJMc2029240}.

\bibitem{Swamy_MBEC24_PulseOximeter}
S.~K.~N. Swamy, C.~He, B.~R. {Hayes-Gill}, {\em et~al.}, ``Pulse oximeter bench
  tests under different simulated skin tones,'' {\em Medical \& Biological
  Engineering \& Computing}   (2024).
\newblock \url{https://doi.org/10.1007/s11517-024-03091-2}.

\bibitem{Chatterjee_Sensors19_MonteCarlo}
S.~Chatterjee and P.~A. Kyriacou, ``Monte {{Carlo Analysis}} of {{Optical
  Interactions}} in {{Reflectance}} and {{Transmittance Finger
  Photoplethysmography}},'' {\em Sensors} {\bf 19}, 789  (2019).
\newblock \url{https://www.mdpi.com/1424-8220/19/4/789}.

\bibitem{Blaney_App.Sci.22_MethodMeasuring}
G.~Blaney, A.~Sassaroli, and S.~Fantini, ``Method for {{Measuring Absolute
  Optical Properties}} of {{Turbid Samples}} in a {{Standard Cuvette}},'' {\em
  Applied Sciences} {\bf 12}, 10903  (2022).
\newblock \url{doi.org/10/gsgpzq}.

\bibitem{Blaney_JBO23_DualratioApproach}
G.~Blaney, F.~Ivich, A.~Sassaroli, {\em et~al.}, ``Dual-ratio approach for
  detection of point fluorophores in biological tissue,'' {\em Journal of
  Biomedical Optics} {\bf 28}, 077001  (2023).
\newblock
  \url{https://www.spiedigitallibrary.org/journals/journal-of-biomedical-optics/volume-28/issue-7/077001/Dual-ratio-approach-for-detection-of-point-fluorophores-in-biological/10.1117/1.JBO.28.7.077001.full}.

\bibitem{Blaney_JBio20_PhaseDualslopes}
G.~Blaney, A.~Sassaroli, T.~Pham, {\em et~al.}, ``Phase dual-slopes in
  frequency-domain near-infrared spectroscopy for enhanced sensitivity to brain
  tissue: {{First}} applications to human subjects,'' {\em Journal of
  Biophotonics} {\bf 13}, e201960018  (2020).
\newblock \url{https://doi.org/10.1002/jbio.201960018}.

\bibitem{Sassaroli_JOSAA19_DualslopeMethod}
A.~Sassaroli, G.~Blaney, and S.~Fantini, ``Dual-slope method for enhanced depth
  sensitivity in diffuse optical spectroscopy,'' {\em Journal of the Optical
  Society of America A} {\bf 36}, 1743--1761  (2019).
\newblock
  \url{https://www.osapublishing.org/abstract.cfm?URI=josaa-36-10-1743}.

\bibitem{Hueber_Opt.Tomogr.Spectrosc.TissueIII99_NewOptical}
D.~M. Hueber, S.~Fantini, A.~E. Cerussi, {\em et~al.}, ``New optical probe
  designs for absolute (self-calibrating) {{NIR}} tissue hemoglobin
  measurements,'' in {\em Optical {{Tomography}} and {{Spectroscopy}} of
  {{Tissue III}}},   {\bf 3597}, 618--631, SPIE  (1999).
\newblock \url{doi.org/10/d8vhhm}.

\bibitem{Prahl_98_TabulatedMolar}
S.~Prahl, ``Tabulated {{Molar Extinction Coefficient}} for {{Hemoglobin}} in
  {{Water}}.'' \url{https://omlc.org/spectra/hemoglobin/summary.html}  (1998).

\bibitem{Fang_OE09_MonteCarlo}
Q.~Fang and D.~A. Boas, ``Monte {{Carlo Simulation}} of {{Photon Migration}} in
  {{3D Turbid Media Accelerated}} by {{Graphics Processing Units}},'' {\em
  Optics Express} {\bf 17}, 20178--20190  (2009).
\newblock
  \url{https://www.osapublishing.org/oe/abstract.cfm?uri=oe-17-22-20178}.

\bibitem{Jacques_PMB13_OpticalProperties}
S.~L. Jacques, ``Optical properties of biological tissues: A review,'' {\em
  Physics in Medicine and Biology} {\bf 58}, R37--R61  (2013).
\newblock \url{https://doi.org/10.1088/0031-9155/58/11/r37}.

\bibitem{Jacques_PMB13_ErratumOptical}
S.~L. Jacques, ``Erratum: {{Optical}} properties of biological tissues: {{A}}
  review ({{Physics}} in {{Medicine}} and {{Biology}} (2013) 58),'' {\em
  Physics in Medicine and Biology} {\bf 58}(14), 5007--5008  (2013).

\bibitem{Salomatina_JBO06_OpticalProperties}
E.~V. Salomatina, B.~Jiang, J.~Novak, {\em et~al.}, ``Optical properties of
  normal and cancerous human skin in the visible and near-infrared spectral
  range,'' {\em Journal of Biomedical Optics} {\bf 11}, 064026  (2006).
\newblock
  \url{https://www.spiedigitallibrary.org/journals/journal-of-biomedical-optics/volume-11/issue-6/064026/Optical-properties-of-normal-and-cancerous-human-skin-in-the/10.1117/1.2398928.full}.

\bibitem{Ma_OL05_BulkOptical}
X.~Ma, J.~Q. Lu, H.~Ding, {\em et~al.}, ``Bulk optical parameters of porcine
  skin dermis at eight wavelengths from 325 to 1557 nm,'' {\em Optics Letters}
  {\bf 30}, 412--414  (2005).
\newblock \url{https://opg.optica.org/ol/abstract.cfm?uri=ol-30-4-412}.

\bibitem{Warner_JID88_ElectronProbe}
R.~R. Warner, M.~C. Myers, and D.~A. Taylor, ``Electron {{Probe Analysis}} of
  {{Human Skin}}: {{Determination}} of the {{Water Concentration Profile}},''
  {\em Journal of Investigative Dermatology} {\bf 90}, 218--224  (1988).
\newblock
  \url{https://www.sciencedirect.com/science/article/pii/S0022202X88911463}.

\bibitem{Schiebener_JPCRef.Dat.90_RefractiveIndex}
P.~Schiebener, J.~Straub, J.~M.~H. Levelt~Sengers, {\em et~al.}, ``Refractive
  index of water and steam as function of wavelength, temperature and
  density,'' {\em Journal of Physical and Chemical Reference Data} {\bf 19},
  677--717  (1990).
\newblock \url{https://doi.org/10.1063/1.555859}.

\bibitem{Choudhury_PhotonicTher.Diagn.VI10_LinkingVisual}
N.~Choudhury, R.~Samatham, and S.~L. Jacques, ``Linking visual appearance of
  skin to the underlying optical properties using multispectral imaging,'' in
  {\em Photonic {{Therapeutics}} and {{Diagnostics VI}}},   {\bf 7548},
  111--116, SPIE  (2010).
\newblock
  \url{https://www.spiedigitallibrary.org/conference-proceedings-of-spie/7548/75480G/Linking-visual-appearance-of-skin-to-the-underlying-optical-properties/10.1117/12.842648.full}.

\bibitem{Peters_PMB90_OpticalProperties}
V.~G. Peters, D.~R. Wyman, M.~S. Patterson, {\em et~al.}, ``Optical properties
  of normal and diseased human breast tissues in the visible and near
  infrared,'' {\em Physics in Medicine \& Biology} {\bf 35}, 1317  (1990).
\newblock \url{https://dx.doi.org/10.1088/0031-9155/35/9/010}.

\bibitem{Jakubowski_JBO04_MonitoringNeoadjuvant}
D.~B. Jakubowski, A.~E. Cerussi, F.~P. Bevilacqua, {\em et~al.}, ``Monitoring
  neoadjuvant chemotherapy in breast cancer using quantitative diffuse optical
  spectroscopy: A case study,'' {\em Journal of Biomedical Optics} {\bf 9},
  230--238  (2004).
\newblock
  \url{https://www.spiedigitallibrary.org/journals/journal-of-biomedical-optics/volume-9/issue-1/0000/Monitoring-neoadjuvant-chemotherapy-in-breast-cancer-using-quantitative-diffuse-optical/10.1117/1.1629681.full}.

\bibitem{Venkata_12_DeterminationOptical}
R.~S. Venkata, {\em Determination of Optical Scattering Properties of Tissues
  Using Reflectance-Mode Confocal Microscopy}.
\newblock PhD thesis, Oregon Health \& Science University  (2012).
\newblock \url{https://doi.org/10.6083/M4R78C7S}.

\bibitem{Matcher_AO97_VivoMeasurements}
S.~J. Matcher, M.~Cope, and D.~T. Delpy, ``In vivo measurements of the
  wavelength dependence of tissue-scattering coefficients between 760 and 900
  nm measured with time-resolved spectroscopy,'' {\em Applied Optics} {\bf 36},
  386--396  (1997).
\newblock \url{https://opg.optica.org/ao/abstract.cfm?uri=ao-36-1-386}.

\bibitem{Mitchell_JBC45_CHEMICALCOMPOSITION}
H.~H. Mitchell, T.~S. Hamilton, F.~R. Steggerda, {\em et~al.}, ``{{THE CHEMICAL
  COMPOSITION OF THE ADULT HUMAN BODY AND ITS BEARING ON THE BIOCHEMISTRY OF
  GROWTH}},'' {\em Journal of Biological Chemistry} {\bf 158}, 625--637
  (1945).
\newblock
  \url{https://www.sciencedirect.com/science/article/pii/S0021925819513394}.

\bibitem{Bevilacqua_AO00_BroadbandAbsorption}
F.~Bevilacqua, A.~J. Berger, A.~E. Cerussi, {\em et~al.}, ``Broadband
  absorption spectroscopy in turbid media by combined frequency-domain and
  steady-state methods,'' {\em Applied Optics} {\bf 39}, 6498--6507  (2000).
\newblock \url{doi.org/10/fkk8t3}.

\bibitem{Firbank_PMB93_MeasurementOptical}
M.~Firbank, M.~Hiraoka, M.~Essenpreis, {\em et~al.}, ``Measurement of the
  optical properties of the skull in the wavelength range 650-950 nm,'' {\em
  Physics in Medicine \& Biology} {\bf 38}, 503  (1993).
\newblock \url{https://dx.doi.org/10.1088/0031-9155/38/4/002}.

\bibitem{Alexandrakis_PMB05_TomographicBioluminescence}
G.~Alexandrakis, F.~R. Rannou, and A.~F. Chatziioannou, ``Tomographic
  bioluminescence imaging by use of a combined optical-{{PET}} ({{OPET}})
  system: A computer simulation feasibility study,'' {\em Physics in Medicine
  \& Biology} {\bf 50}, 4225  (2005).
\newblock \url{https://dx.doi.org/10.1088/0031-9155/50/17/021}.

\bibitem{Weaver_CBPA89_TissueBlood}
B.~M.~Q. Weaver, G.~E. Staddon, and M.~R.~B. Pearson, ``Tissue blood content in
  anaesthetised sheep and horses,'' {\em Comparative Biochemistry and
  Physiology Part A: Physiology} {\bf 94}, 401--404  (1989).
\newblock
  \url{https://www.sciencedirect.com/science/article/pii/0300962989901138}.

\bibitem{Harrison_Blood02_OxygenSaturation}
J.~S. Harrison, P.~Rameshwar, V.~Chang, {\em et~al.}, ``Oxygen saturation in
  the bone marrow of healthy volunteers,'' {\em Blood} {\bf 99}, 394  (2002).
\newblock \url{https://doi.org/10.1182/blood.V99.1.394}.

\bibitem{Jacques_98_MelanosomeAbsorption}
S.~Jacques, ``Melanosome {{Absorption Coefficient}}.''
  \url{https://omlc.org/spectra/melanin/mua.html}  (1998).

\bibitem{Jacques_18_ExtinctionCoefficient}
S.~Jacques, ``Extinction {{Coefficient}} of {{Melanin}}.''
  \url{https://omlc.org/spectra/melanin/extcoeff.html}  (2018).

\bibitem{Jacques_18_OpticalAbsorption}
S.~Jacques, ``Optical {{Absorption}} of {{Melanin}}.''
  \url{https://omlc.org/spectra/melanin/index.html}  (2018).

\bibitem{Jacques_PP91_MelanosomeThreshold}
S.~L. Jacques and D.~J. McAuliffe, ``The {{Melanosome}}: {{Threshold
  Temperature}} for {{Explosive Vaporization}} and {{Internal Absorption
  Coefficient During Pulsed Laser Irradiation}},'' {\em Photochemistry and
  Photobiology} {\bf 53}(6), 769--775  (1991).
\newblock
  \url{https://onlinelibrary.wiley.com/doi/abs/10.1111/j.1751-1097.1991.tb09891.x}.

\bibitem{Blaney_JIOHS24_SpatialSensitivity}
G.~Blaney, A.~Sassaroli, and S.~Fantini, ``Spatial sensitivity to absorption
  changes for various near-infrared spectroscopy methods: {{A}} compendium
  review,'' {\em Journal of Innovative Optical Health Sciences} , 2430001
  (2024).
\newblock \url{https://www.worldscientific.com/doi/10.1142/S1793545824300015}.

\bibitem{Blaney_22_EnablingDeep}
G.~Blaney, {\em Enabling {{Deep Region Specific Optical Measurements}} in a
  {{Diffusive Medium}} with {{Near-Infrared Spectroscopy}}}.
\newblock PhD thesis, Tufts University, Medford, MA USA  (2022).
\newblock
  \url{proquest.com/dissertations-theses/enabling-deep-region-specific-optical/docview/2676526855/se-2}.

\bibitem{Yao_BOE18_DirectApproach}
R.~Yao, X.~Intes, and Q.~Fang, ``Direct approach to compute {{Jacobians}} for
  diffuse optical tomography using perturbation {{Monte Carlo-based}} photon
  ``replay'','' {\em Biomedical Optics Express} {\bf 9}, 4588--4603  (2018).
\newblock \url{https://opg.optica.org/boe/abstract.cfm?uri=boe-9-10-4588}.

\bibitem{Aronson_JOSAA97_RadiativeTransfer}
R.~Aronson, ``Radiative transfer implies a modified reciprocity relation,''
  {\em JOSA A} {\bf 14}, 486--490  (1997).
\newblock \url{https://opg.optica.org/josaa/abstract.cfm?uri=josaa-14-2-486}.

\bibitem{Fantini_NeuroImage14_DynamicModel}
S.~Fantini, ``Dynamic model for the tissue concentration and oxygen saturation
  of hemoglobin in relation to blood volume, flow velocity, and oxygen
  consumption: {{Implications}} for functional neuroimaging and coherent
  hemodynamics spectroscopy ({{CHS}}),'' {\em NeuroImage} {\bf 85}, 202--221
  (2014).
\newblock \url{http://dx.doi.org/10.1016/j.neuroimage.2013.03.065}.

\bibitem{Kainerstorfer_JBO16_OpticalOximetry}
J.~M. Kainerstorfer, A.~Sassaroli, and S.~Fantini, ``Optical oximetry of
  volume-oscillating vascular compartments: Contributions from oscillatory
  blood flow,'' {\em Journal of Biomedical Optics} {\bf 21}, 101408  (2016).
\newblock
  \url{https://www.spiedigitallibrary.org/journals/journal-of-biomedical-optics/volume-21/issue-10/101408/Optical-oximetry-of-volume-oscillating-vascular-compartments--contributions-from/10.1117/1.JBO.21.10.101408.full}.

\bibitem{Nuttall_IEEETrans.81_WindowsVery}
A.~Nuttall, ``Some windows with very good sidelobe behavior,'' {\em IEEE
  Transactions on Acoustics, Speech, and Signal Processing} {\bf 29}, 84--91
  (1981).
\newblock \url{https://ieeexplore.ieee.org/document/1163506}.

\bibitem{Monk_2019}
E.~Monk, ``Monk skin tone scale.'' \url{https://skintone.google}  (2019).

\end{thebibliography}
\bibliographystyle{spiejour}   

\vspace{2ex}\noindent\textbf{Giles~Blaney} is a National Institutes of Health (NIH) Institutional Research and Academic Career Development Award (IRACDA) Postdoctoral Scholar in the Diffuse Optical Imaging of Tissue (DOIT) lab at Tufts University. He received his Ph.D. from Tufts University (Medford, MA USA) in 2022 after working in the same lab with Prof.~Sergio~Fantini as his advisor. Before that Giles received an undergraduate degree in Mechanical Engineering and Physics from Northeastern University (Boston, MA USA). His current research interests include diffuse optics and its possible applications within and outside of medical imaging.

\vspace{2ex}\noindent\textbf{Jodee~Frias} is a second-year Ph.D. student in the DOIT Lab at Tufts University. She is from Avon, MA and received a Bachelor of Science in Biomedical Engineering from Boston University in 2022. During her undergraduate degree, she conducted research on a wearable Short-Wave InfraRed (SWIR) optical probe to monitor hydration in hemodialysis patients. Jodee is currently working on functional Near-InfraRed Spectroscopy (NIRS) measurements, and is extremely interested in non-invasive imaging techniques for clinical applications.

\vspace{2ex}\noindent\textbf{Fatemeh~Tavakoli} is a second-year Ph.D. student in the Diffuse Optical Imaging of Tissue (DOIT) lab under the advisement of Professor Sergio Fantini at Tufts University (Medford, Massachusetts). Her current field of research is Frequency-Domain Near-Infrared Spectroscopy (NIRS) and Diffuse Optics to investigate non-invasive techniques for medical applications, such as hemodynamic monitoring of skeletal muscles and brain function. Before that, she received an M.Sc. degree in Electrical Engineering from Islamic Azad University, Science and Research Branch, in 2020 (Tehran, Iran). During her master's degree, she found Fractal geometry to be a unique way of looking at the world, and she successfully designed a perfect absorber based on self-similar nanoparticles, allowing for tuning optical properties for near-infrared applications. 

\vspace{2ex}\noindent\textbf{Angelo~Sassaroli} received a Ph.D. in Physics in 2002 from the University of Electro-Communications (Tokyo, Japan). From July~2002 to August~2007, he was a Research Associate in the research group of Prof.~Sergio~Fantini at Tufts University. In September~2007 he was appointed by Tufts University as a Research Assistant Professor. His field of research is near-infrared spectroscopy and diffuse optical tomography.

\vspace{2ex}\noindent\textbf{Sergio~Fantini} is a Professor of Biomedical Engineering and Principal Investigator of the DOIT at Tufts University. His research activities on applying diffuse optics to biological tissues resulted in about \num{120} peer-reviewed scientific publications and \num{12} patents. He co-authored with Prof.~Irving~Bigio (Boston University, Boston, MA USA) a textbook on \say{Quantitative Biomedical Optics} published by Cambridge University Press in 2016. He is a Fellow of SPIE, Optica, and the American Institute for Medical and Biological Engineering (AIMBE).



\end{spacing}
\end{document}